\documentclass[twocolumn,superscriptaddress,prb,aps,preprintnumbers,nofootinbib,longbibliography]{revtex4-2}
\usepackage{amsmath,amssymb,graphicx,color}
\usepackage[colorlinks=true, citecolor={blue!80!black}, urlcolor={blue!50!black}, linkcolor = {blue!80!black}]{hyperref}
\usepackage{comment}

\usepackage{xcolor}

\newcommand{\COHB}{Cu$_2$(OH)$_3$Br}
\newcommand{\COHC}{Cu$_2$(OH)$_3$Cl}

\newcommand{\hamilt}{\hat{\mathcal{H}}}
\newcommand{\spin}{\hat{S}}
\newcommand{\spop}{\hat{\mathbf{S}}}

\newcommand{\aver}[1]{\left\langle #1 \right\rangle}

\newcommand{\mub}{\mu_\mathrm{B}}
\newcommand{\kb}{k_\mathrm{B}}

\begin{document}
\title{ High-field magnetic properties of the alternating ferro-antiferromagnetic spin-chain compound ~Cu$_2$(OH)$_3$Br}

\author{K. Yu. Povarov} 
\email{k.povarov@hzdr.de}
\affiliation{Dresden High Magnetic Field Laboratory (HLD-EMFL) and W\"urzburg-Dresden Cluster of Excellence ct.qmat, Helmholtz-Zentrum Dresden-Rossendorf (HZDR), 01328 Dresden, Germany}

\author{Y.~Skourskii}
\affiliation{Dresden High Magnetic Field Laboratory (HLD-EMFL) and W\"urzburg-Dresden Cluster of Excellence ct.qmat, Helmholtz-Zentrum Dresden-Rossendorf (HZDR), 01328 Dresden, Germany}

\author{J.~Wosnitza} 
\affiliation{Dresden High Magnetic Field Laboratory (HLD-EMFL) and W\"urzburg-Dresden Cluster of Excellence ct.qmat, Helmholtz-Zentrum Dresden-Rossendorf (HZDR), 01328 Dresden, Germany}
\affiliation{Institut f\"ur Festk\"orper- und Materialphysik, Technische Universit\"at Dresden, 01062 Dresden, Germany}

\author{D. E. Graf}
\affiliation{National High Magnetic Field Laboratory, Tallahassee, Florida 32310, USA}

\author{Z. Zhao}
\email{zhiyingzhao@fjirsm.ac.cn}
\affiliation{State Key Laboratory of Structural Chemistry, Fujian Institute of Research on the Structure of Matter,
Chinese Academy of Sciences, Fuzhou, Fujian 350002, People’s Republic of China}

\author{S.~A.~Zvyagin} 
\email{s.zvyagin@hzdr.de}
\affiliation{Dresden High Magnetic Field Laboratory (HLD-EMFL) and W\"urzburg-Dresden Cluster of Excellence ct.qmat, Helmholtz-Zentrum Dresden-Rossendorf (HZDR), 01328 Dresden, Germany}

\begin{abstract}
We present comprehensive high magnetic field studies of the alternating weakly coupled ferro-antiferromagnetic (FM-AFM) spin-$1/2$ chain compound Cu$_2$(OH)$_3$Br, with the structure of the natural mineral botallackite. Our measurements reveal a broad magnetization plateau at about half of the saturation value, strongly suggesting that the FM chain sublattice becomes fully polarized, while the AFM chain sublattice remains barely magnetized, in magnetic fields at least up to $50$~T. We confirm a spin-reorientation transition for magnetic fields applied in the $ac^\ast$-plane, whose angular dependence is described in the framework of the mean-field theory. Employing high-field THz spectroscopy, we reveal a complex pattern of high-frequency spinon-magnon bound-state excitations. On the other hand, at lower frequencies we observe two modes of antiferromagnetic resonance, as a consequence of the long-range magnetic ordering. We demonstrate that applied magnetic field tends to suppress the long-range magnetic ordering; the temperature-field phase diagram of the phase transition is obtained for magnetic fields up to $14$~T for three principal directions ($a$, $b$, $c^\ast$).
\end{abstract}
\date{\today}
\maketitle

\section{Introduction}
Natural minerals featuring triangular motifs in $S=1/2$ Cu$^{2+}$ ion arrangements are attracting significant attention ~\cite{Inosov_AdvPhys_2018_MineralReview}. Prominent examples include the frustrated spin-chain linarite~\cite{Willenberg_PRL_2012_LinariteFrustrated,Willenberg_PRL_2016_LinariteSDWs,PovarovFeng_PRB_2016_LinariteMF,CemalEnderle_PRL_2018_Linarite}, the spin-liquid candidate herbertsmithite~\cite{MendelsBert_PRL_2007_HerbertNoLRO,Norman_RMP_2016_HerbReview,BarthelemyDemuer_PRX_2022_HerbertsmCp}, the delta-chain material atacamite~\cite{HeinzeJeshke_PRL_2021_AtacamitePlateau}, and the three-leg ladder system antlerite~\cite{KulbakovKononenko_PRB_2022_Antlerite,KulbakovSadrollahi_PRB_2022_AntleriteQ}, to name a few. Since the triangular-like geometry of magnetic correlations implies competing interactions and the frustration~\cite{Diep_2013_FrustBook,Starykh_RepPrPhys_2015_TriangularReview}, such materials often demonstrate diverse and complex magnetic properties.

\COHB\ (hereafter COHB) is a sister compound of the natural mineral botallackite \COHC, which has been recently identified as a new spin-$1/2$ chain quantum system with a triangular motif of exchange couplings~\cite{ZhaoChe_JPCM_2019_COHB,ZhangZhao_PRL_2020_COHBspectrum,Gautreau_PRM_2021_COHBdft}.
Employing inelastic neutron scattering, Zhang~\emph{et al.} ~\cite{ZhangZhao_PRL_2020_COHBspectrum} has established the presence of alternating and weakly coupled ferromagnetic (FM) and antiferromagnetic (AFM) chains as the key feature of the COHB magnetic structure. Remarkably, the neutron scattering has revealed signatures of the co-existing two-spinon continuum and FM magnons (as well as magnon-spinon bound states resulting from the interaction between FM and AFM chain sublattices)~\cite{ZhangZhao_PRL_2020_COHBspectrum}. It was suggested that frustrated interactions between the chains of different types is the specific feature of COHB, leading to intricate ground state and magnetic excitations.

It is worthwhile to mention that not much is known about high-magnetic-field properties of COHB, including its magnetic phase diagram and the high-field spin dynamics. The most important finding in this regards is the field-induced spin-reorientation phase transition in the magnetically ordered state below $T_\mathrm{N}=9.3$~K, with magnetic field applied
in the $ac$ plane and $\mu_{0}H_\mathrm{c}$ between approximately 4.5 and 5~T (for $a$ and $c^\ast$ directions)~\cite{ZhaoChe_JPCM_2019_COHB,XiaoOuyang_JPCM_2022_COHBangular}. It was shown by the electron spin resonance (ESR spectroscopy) that this this discontinuous phase transition is accompanied by a partial softening of a gapped resonance mode~\cite{XiaoOuyang_JPCM_2022_COHBangular}. The exact nature of the transition remains an open question.
In this work, employing high-field magnetization, tunnel-diode-oscillator (TDO) susceptibility, and ESR spectroscopy techniques, we study magnetic properties of COHB. This allows us to reveal several important features of its high-field behavior, originating from the interplay between the FM and AFM chain sublattices.

\begin{figure}
\includegraphics[width=0.48\textwidth]{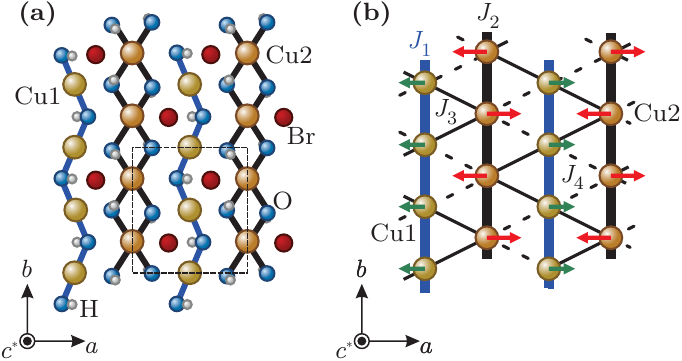}
\caption{(a) Schematic view of the crystal structure of COHB. Orange spheres represent the Cu$^{2+}$ ions; blue, red, and grey ones stand for O, Br, and H atoms, correspondingly. The dashed line shows the unit cell. Cu$1$ ions form ferromagnetic chains, and Cu$2$ ions form antiferromagnetic chains. (b) Schematic view of the magnetic structure of COHB. The projections of the Cu spins in the $ab$ plane (following~\cite{ZhangZhao_PRL_2020_COHBspectrum}) are indicated by arrows. Thick blue lines, thick black lines, thin solid lines, and thin dashed lines correspond to exchange interactions $J_1$, $J_2$, $J_3$, and $J_4$ respectively.}\label{FIG:xtal}
\end{figure}

\section{Crystal and magnetic structure}

COHB crystallizes in the monoclinic structure with space group $P2_{1}/m$ and lattice parameters $a=5.63$~\AA, $b=6.12$~\AA, $c=5.72$~\AA, $\beta=93.1^{\circ}$, and $Z=4$~\cite{ZhengMori_PRB_2005_COHXstructure,Krivovichev_StrChem_2017_COHXstructure}. A schematic view of the COHB crystal structure is shown in Fig.~\ref{FIG:xtal}(a). There are two
types of spin chains, with sites Cu$1$ and Cu$2$, respectively. The chains are arranged in well-separated layers in the $ab$ plane, running
along the $b$ direction.

Neutron diffraction in COHB below $T_\mathrm{N}=9.3$~K revealed a peculiar spin order with the propagation vector $\mathbf{Q}=(1/2,0,0)$~\cite{ZhangZhao_PRL_2020_COHBspectrum}. It was shown that along the $b$ axis, Cu$1$ spins are aligned
ferromagnetically with spins oriented nearly along the
diagonal direction in the $ac$ plane, while Cu$2$ spins are aligned
antiferromagnetically with spins oriented along the $a$ axis [Fig.~\ref{FIG:xtal}(b)].

As follows from Ref.~\cite{Gautreau_PRM_2021_COHBdft}, the intrachain exchange interactions are  $J_{1}/\kb=-16.2$~K and $J_{2}/\kb=55.7$~K (for FM and AFM chains, respectively), while the interchain exchange interactions are  $J_{3}/\kb=9.3$~K and $J_{4}/\kb=4.6$~K [see Fig.~\ref{FIG:xtal}(b)]. These values are in line with estimates given in Ref.~\cite{ZhangZhao_PRL_2020_COHBspectrum}.

\section{Experimental}
Single crystals of COHB were grown using a conventional hydrothermal method, as described in~Ref. \cite{ZhaoChe_JPCM_2019_COHB}. We measured the magnetization in DC fields up to 14 T using a vibrating sample magnetometer (VSM) (product of Quantum Design, Inc.). For pulsed-field magnetization measurements, we used a coaxial pick-up coil magnetometer (PUCM), similar to that described in Ref.~\cite{SkourskiKuzmin_PRB_2011_PulseM}. We measured the high-frequency susceptibility in DC fields up to 41.5 T using a TDO magnetometer \cite{CloverWulf_RevSciInstr_1970_TDO,GhannadzadehMoeller_PRB_2013_earlyTDO,ShiDissanayakeCorboz_NatComm_2022_SCBOpressTDO}. In our ESR experiments, we employed a THz-range spectrometer (similar to the one described in Ref.~\cite{Zvyagin_PhysB_2004_ESRinCuGeO3}), equipped with a $16$~T superconducting magnet. We used VDI microwave-chain sources (product of Virginia Diodes, Inc., USA) and backward wave oscillators (product of NPP Istok, Russian Federation) to generate radiation in the frequency range of $50-900$~GHz. A hot-electron n-InSb bolometer (product of QMC Instruments Ltd., UK), operated at 4.2 K, was employed as a THz detector. For the ESR experiments we used a probe in the Voigt geometry for $H\parallel a$ and $b$, while for $H\parallel c^\ast$ we used the Faraday geometry. We used 2,2-diphenyl-1-picrylhydrazyl (DPPH) as a standard frequency-field marker. The crystals were oriented with $\pm5^\circ$ accuracy.

\begin{figure}
\includegraphics[width=0.48\textwidth]{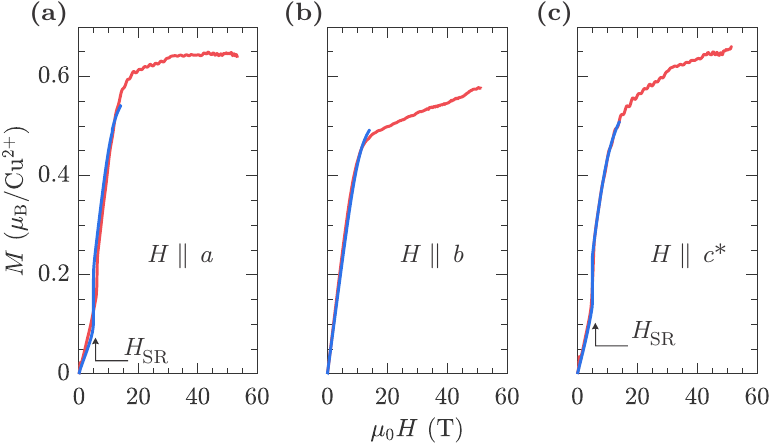}
\caption{(a)-(c) Magnetization of COHB single crystals for magnetic fields applied along the $a,b$ and $c^*$ directions, respectively. Blue lines are the data, obtained at $T=2$~K using VSM up to $14$~T; red lines correspond to pulsed-field magnetization data at $T=1.5$~K, obtained employing PUCM up to $55$~T. The arrows mark the spin-reorientation transition field $H_\mathrm{SR}$.}\label{FIG:M}
\end{figure}

\section{High-field magnetization}

In Fig.~\ref{FIG:M}, we present magnetization results for $H\parallel a,b,c^\ast$. For $H\parallel a$, $c^\ast$, we observe magnetization steps at $\mu_{0}H_\mathrm{SR}\simeq4.9$~T and $5$~T, respectively. These discontinuities corresponds to the spin-reorientation transition, which was previously reported by Zhao~\emph{et al.}~\cite{ZhaoChe_JPCM_2019_COHB}. No steps appear for $H\parallel b$. In addition, at higher fields, we observe a broad plateau at half of the nominal full magnetization (that would be close to one Bohr magneton $\mub$ per Cu$^{2+}$), suggesting that half of the Cu$^{2+}$ ions are polarized. This effective $1/2$-magnetization plateau begins to develop around $10-15$~T and is visible up to $55$~T. Since the mutual orientation of FM chain sublattices appears to be defined by smaller exchange couplings $J_{3,4}/\kb< 10$~K, it is evident that the magnetic field drives the FM chain sublattice to the full saturation first, leaving, on the other hand, the AFM chain sublattice with $J_{2}/\kb\simeq 60$~K only barely magnetized in this field range.

\section{Magnetic phase diagram}

\begin{figure}
\includegraphics[width=0.48\textwidth]{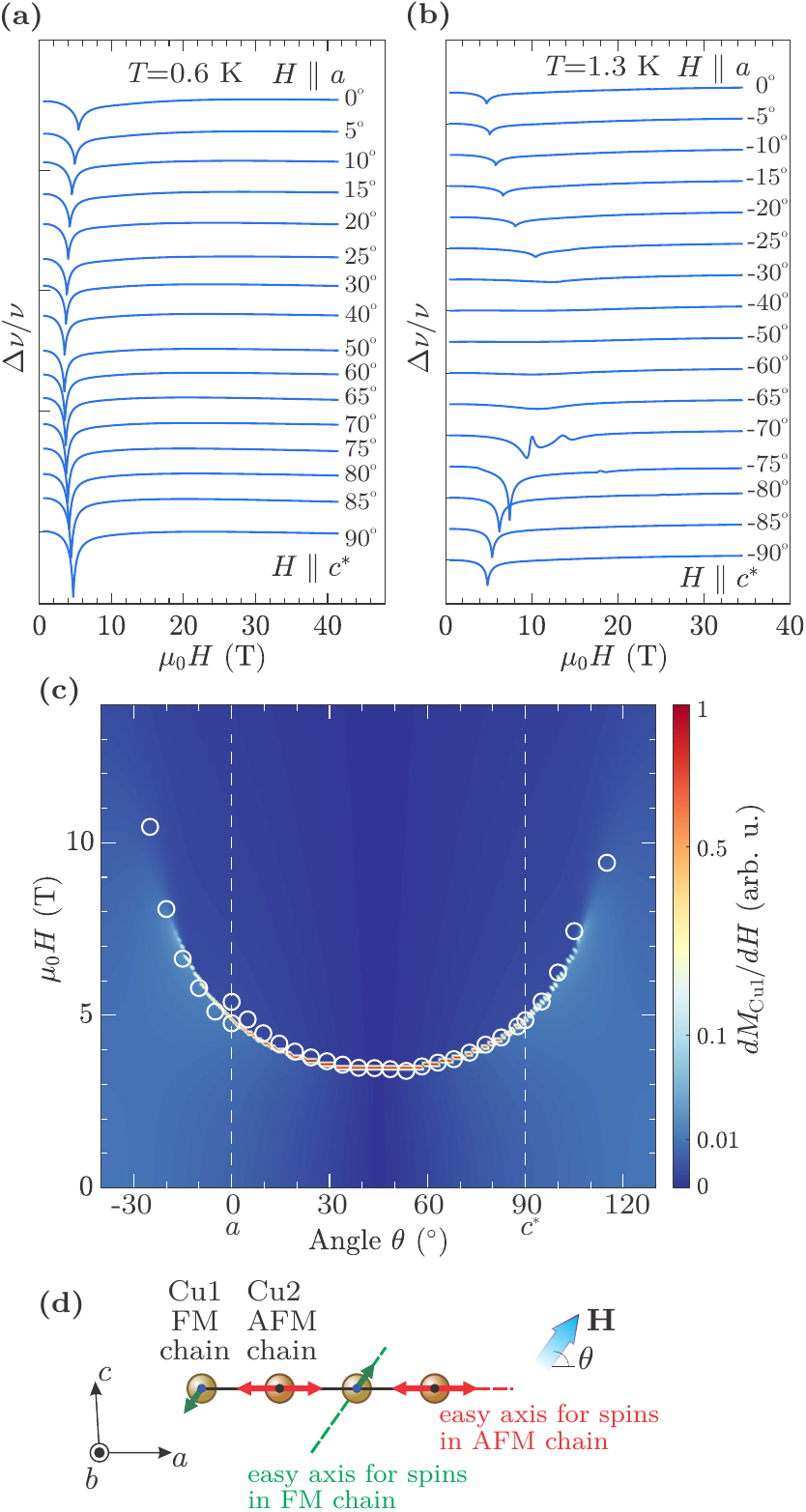}
\caption{
(a) Relative changes of the TDO-circuit frequency for fields applied between $a$ ($\theta=0^\circ$) and $c^\ast$ ($\theta=90^\circ$) at $0.6$~K. (b) Relative change of TDO-circuit frequency for fields applied between $a$ ($\theta=0^\circ$) and $-c^\ast$ ($\theta=-90^\circ$) at $1.3$~K.
(c) Angular dependence of $H_\mathrm{SR}$ as determined from the TDO anomalies (circles).
For comparison, we show the calculated Cu$1$ chain sublattice spin susceptibility (colormap; see Appendix~\ref{APP:SF}). (d) Intrachain anisotropy axes and zero-field spin arrangement in COHB, view along $b$. Small green and red arrows are spin directions at Cu$1$ and Cu$2$ sites, respectively (according to Ref.~\cite{ZhangZhao_PRL_2020_COHBspectrum}). Green and red dashed lines show the directions of the easy axes for spins in the corresponding chains~(after Ref.~\cite{XiaoOuyang_JPCM_2022_COHBangular}). The thick blue arrow represents the external magnetic field.
}\label{FIG:TDO}
\end{figure}

\begin{figure*}[hbt!]
\includegraphics[width=0.9\textwidth]{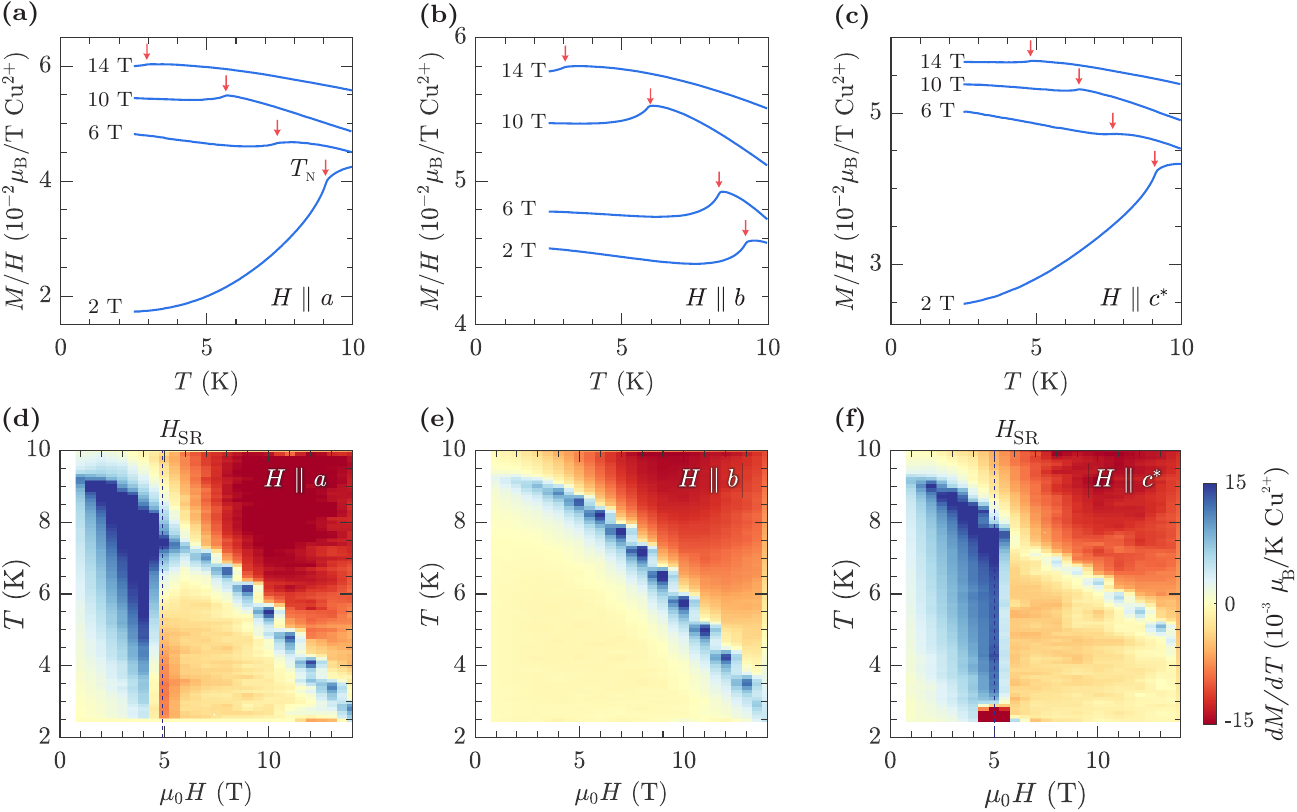}
\caption{(a)-(c) Examples of temperature dependencies of the field-normalized magnetization $M/H$ of COHB at $2$, $6$, $10$, and $14$~T in magnetic fields $H\parallel a,b$, and $c^\ast$, respectively. The arrows indicate anomalies, related to the long-range ordering. The data are offset for clarity.
(d)-(f) Colormap of the COHB magnetization derivative with respect to temperature, $dM/dT$, with the magnetic fields applied along $a,b$, and $c^\ast$, respectively. The magnetic field step is $1$~T. Black dashed lines mark the spin-reorientation transition field.}\label{FIG:VSM}
\end{figure*}

In Figs.~\ref{FIG:TDO}(a)-(c), we show the angular dependence of the SR transition, observed in COHB below $T_\mathrm{N}$. As revealed from our experiments, the spin-reorientation transition is visible not only for $H\parallel a,c^\ast$ directions, but for a wide range of angles in $ac^\ast$ plane.
 The dependence has a minimum; this field direction corresponds to the easy-axis anisotropy direction for spins in the FM chains sublattice [Fig.~\ref{FIG:TDO}(d)], as determined by neutron-diffraction measurements~\cite{ZhangZhao_PRL_2020_COHBspectrum} and magnetization angular dependence~\cite{XiaoOuyang_JPCM_2022_COHBangular}. The present angular dependence is in a good agreement with the one obtained in Ref.~\cite{XiaoOuyang_JPCM_2022_COHBangular} and their phenomenological description.
For a more detailed analysis of the $H_\mathrm{SR}(\theta)$ we also account for interactions between the AFM and FM chain sublattices. In addition to the Heisenberg couplings $J_3$ and $J_4$, we consider possible antisymmetric Dzyaloshinskii--Moriya (DM) interactions~\cite{Dzyaloshinskii_JChemPhysSol_1958_DM,Moriya_PR_1960_DM} between the chains, given as $\hamilt_\mathrm{DM}=\mathbf{D}\cdot[\spop_\mathrm{Cu1}\times\spop_\mathrm{Cu2}]$. Such interactions ($\mathbf{D}_3$ on $J_3$-type bonds, and $\mathbf{D}_4$ on $J_4$-type bonds) are allowed by the symmetry of COHB, with the corresponding DM vector directions following a complex pattern. Neutron spectroscopy suggests that those contributions can be relatively strong~\cite{ZhangZhao_PRL_2020_COHBspectrum}. The DM pattern and the calculation details are described in the Appendices~\ref{APP:DM} and \ref{APP:SF}. Here we provide a brief summary.
The model assumes fixed orientation of AFM spins, while the FM spins experience a combination of intricate mean-field and the external field. The analysis suggests the following form of the molecular field acting on the FM chains:
 \begin{equation}
 \pm g\mub\mu_{0}\mathbf{H}_\mathrm{mol}=(J_{3}-J_{4},~0,~-D_{3b}-D_{4b}).
 \end{equation}
Here the $D_{3b,4b}$ are the components of the corresponding DM vectors along the $b$ axis. The sign of this molecular field is different for neighboring FM chains.
The experimental angular dependence of the spin-reorientation transition field is most correctly reproduced assuming both the easy-axis direction for spins in the FM chains and the field $\mathbf{H}_\mathrm{mol}$ colaigned at nearly $45^\circ$ to the $a$ axis. This is the experimentally observed direction of Cu1 spins at zero field~\cite{ZhangZhao_PRL_2020_COHBspectrum}.
 Figure~\ref{FIG:TDO}(c) shows a comparison of the mean-field model calculations with our experimental TDO data. The calculations yield a $J_{1}$ anisotropy parameter (relative increase in interaction strength along the preferred direction) of $\delta\simeq0.2$, which is in a good agreement with the value $\delta=0.17$ obtained from linear spin-wave theory~\cite{ZhangZhao_PRL_2020_COHBspectrum}. The mean-field model also suggests $(J_{3}-J_{4})/\kb\simeq2.5$~K. In the spin-wave analysis of Ref.~\cite{ZhangZhao_PRL_2020_COHBspectrum} this difference is about $10$~K. Thus, the interchain interactions appear to be more frustrated according to the mean-field estimate. The $b$ components of the DM vectors on these bonds supposedly have a similar magnitude of about $|D_{3b}+D_{4b}|/\kb\simeq2.5$~K.
With these parameters, the FM chains also create a staggered field along $a$ on the AFM sites, providing consistent description of the zero-field magnetic state proposed in Ref.~\cite{ZhangZhao_PRL_2020_COHBspectrum}.

Now we discuss the phase diagram of COHB in fields above $H_\mathrm{SR}$.
In Figs.~\ref{FIG:VSM}(a)-\ref{FIG:VSM}(c), we show the exemplary temperature dependences of the COHB magnetization for fields applied along the $a$, $b$, and $c^\ast$ axes, respectively. The transition into the magnetically ordered phase is well visible as a cusp in the $M(T)$ curves.

In Figs.~\ref{FIG:VSM}(d)-\ref{FIG:VSM}(f), we present a colormap of the temperature derivative of the magnetization $dM/dT$ in magnetic fields $H\parallel a,b$, and $c^\ast$, respectively (field step is $1$~T). The applied magnetic field suppresses the long-range ordered phase.
An asymmetry of the corresponding phase boundary between $c^\ast$ and other field directions is evident from the data present in Fig.~\ref{FIG:VSM}. This is a very unusual and important observation. It is known that in the presence of staggered terms (such as the DM interactions) the application of magnetic field can result in intricate effective staggered fields~\cite{OshikawaAffleck_PRL_1997_staggered}, affecting interchain spin correlations and thus varying $H_c$. Such an effect was observed, e.g.,
in the spin-chain system CuCl$_{2}\cdot 2$[(CD$_{3}$)$_{2}$SO]~\cite{ChenStone_PRB_2007_CDCstaggeredOP}. It is likely that similar effects are responsible for the pronounced anisotropy observed in COHB as well.

\section{Magnetic excitations}

\begin{figure*}
\includegraphics[width=\textwidth]{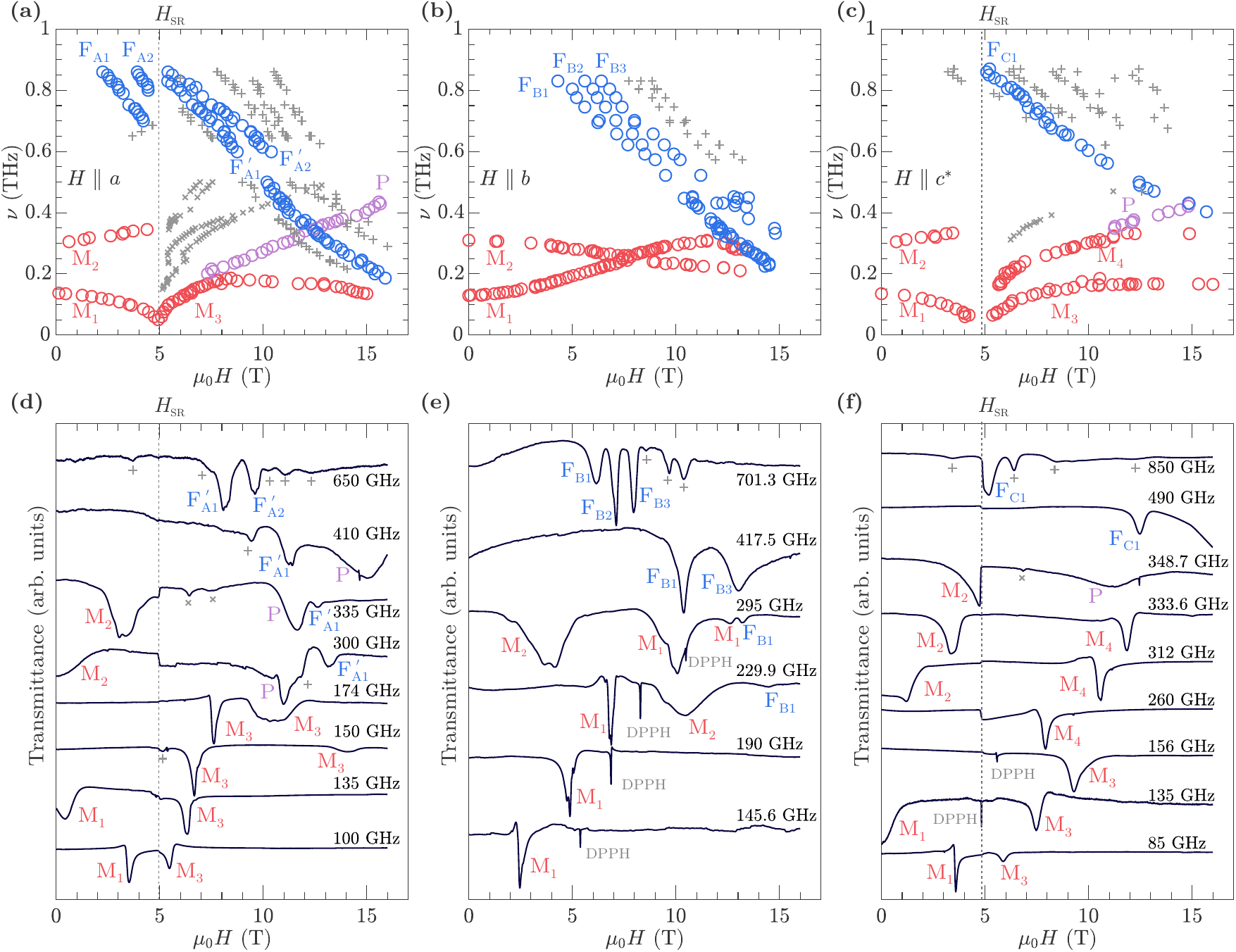}
\caption{(a)-(c) Frequency-field diagrams of the ESR excitations in COHB for $H\parallel a,b$, and $c^\ast$, respectively ($T=1.5$~K). Modes M correspond to AFMR (red circles), while modes F correpond to magnon-spinon bound state excitations (blue circles). The modes P correspond to broad resonances at high fields, that is likely related to spinon deconfinement (purple circles). Crosses denote weaker unclassified resonances. (d)-(f) Selected examples of ESR spectra for $H\parallel a,b$, and $c^\ast$, respectively. The dashed lines mark the spin-reorientation field $H_\mathrm{SR}$. }\label{FIG:six}
\end{figure*}

\begin{figure*}[t]
\includegraphics[width=0.99\textwidth]{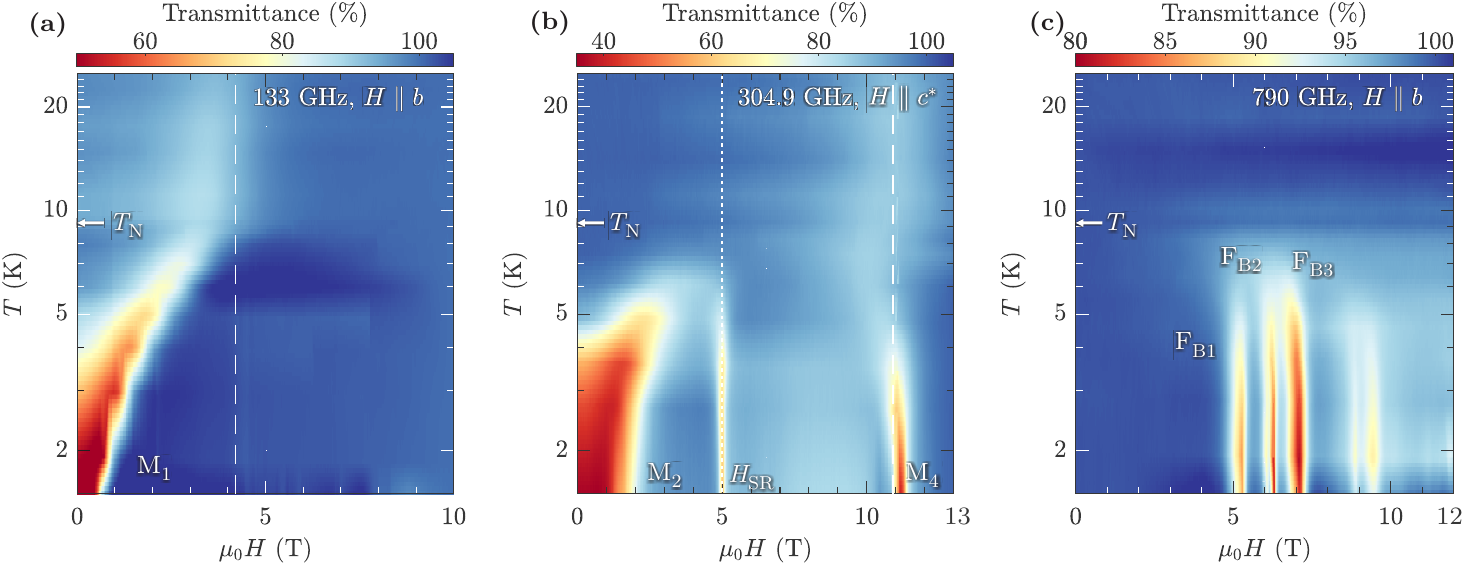}
\caption{THz radiation transmittance vs. field and temperature for $\nu=133$~GHz and $H\parallel b$ (a), $\nu=304.9$~GHz and $H\parallel c^\ast$ (b), as well as for $\nu=790$~GHz and $H\parallel b$ (c). Arrows indicate the zero-field N\'eel temperature, vertical dashed lines show the paramagnetic resonance position above this temperature, and the vertical dotted line marks the $H_\mathrm{SR}$ field. The transmittance is normalized to the respective value far away from the resonance field. Note the logarithmic temperature scale.}\label{FIG:Tdep}
\end{figure*}

\begin{figure}[t]
\includegraphics[width=0.48\textwidth]{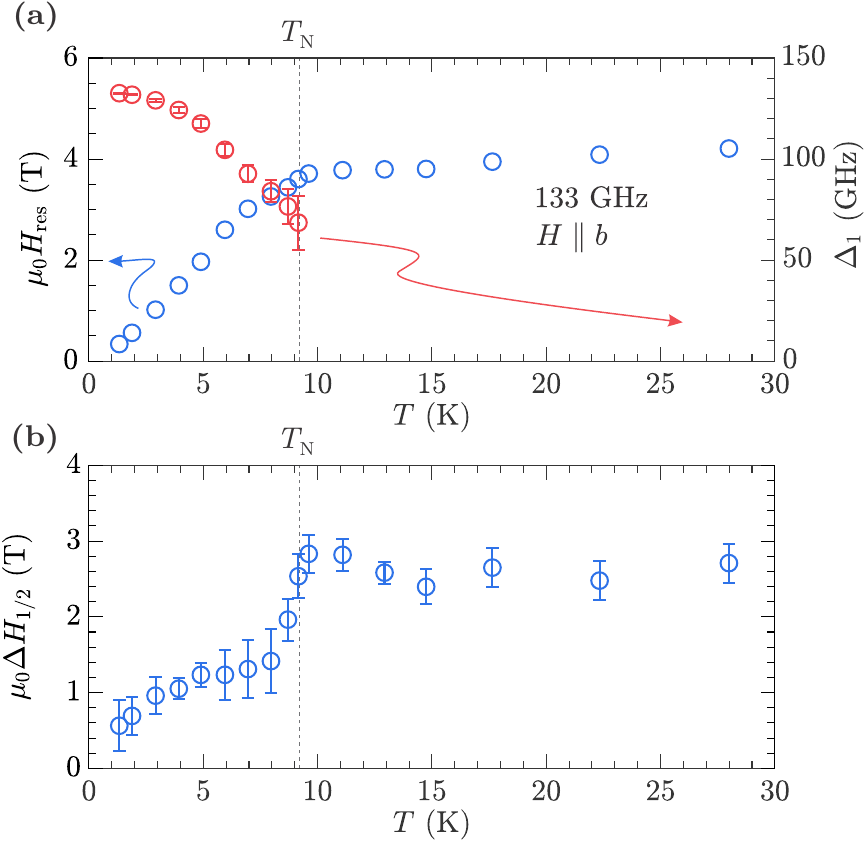}
\caption{ (a) Temperature dependence of the ESR field (blue, right axis) at $133$~GHz, $H\parallel b$, and of the corresponding gap $\Delta_1$ (red, left axis) extracted from the data according to Eq.~(\ref{EQ:AFMR}). (b) ESR linewidth at $133$~GHz, $H\parallel b$. Dotted line marks the N\'eel temperature $T_\mathrm{N}$.}\label{FIG:TdepM1}
\end{figure}

In Figs.~\ref{FIG:six}(a)-\ref{FIG:six}(c) we present frequency-field diagrams of magnetic excitations in COHB for $H\parallel a,b$, and $c^\ast$, respectively.
Corresponding examples of ESR spectra are shown in Figs.~\ref{FIG:six}(d)-\ref{FIG:six}(f). For these three orientations, we observed multiple excitations at relatively high frequencies (labeled as F$_\mathrm{A1,2}$, F$_\mathrm{B1,2,3}$, and F$_\mathrm{C1}$ for $H\parallel a,b$, and $c^\ast$, respectively). The excitation frequencies of these resonances
decrease with the applied magnetic field in a nearly linear fashion. Linear extrapolations of the frequency-field dependencies of these excitations to $H=0$ yield $1$~THz to $1.2$~THz, which agrees with the energy range ($4-5$~meV) for spinon-magnon bound states at the $\Gamma$ point revealed by neutron-scattering experiments~\cite{ZhangZhao_PRL_2020_COHBspectrum}.

Since COHB undergoes a transition into a magnetically ordered state at $T_\mathrm{N}=9.3$ K, below this temperature one would expect the presence of at least two relativistic modes of antiferromagnetic resonance (AFMR). We indeed observed these pseudo-Goldstone modes (low-energy modes M$_1$ and M$_2$). Both AFMR modes are gapped ($\Delta_{1}=130\pm5$~GHz and $\Delta_{2}=300\pm10$~GHz at $T=1.5$~K, respectively), suggesting the presence of biaxial magnetic anisotropy in COHB. Our observation expands the previous low-frequency results of Xiao~\emph{et al}.~\cite{XiaoOuyang_JPCM_2022_COHBangular}, where only the M$_1$ and M$_3$ modes were observed.
The low-frequency AFMR mode M$_1$ exhibits partial softening at $H_\mathrm{SR}$, similar as observed for a conventional spin-flop transition in collinear easy-axis antiferromagnets~\cite{Turov_2004_AFMsbook}. Above the spin-reorientation transition field, we observe AFMR modes M$_3$ and M$_4$ (the latter visible only for $H\parallel c^\ast$).

In addition to the F-labeled multiplets and M-labeled AFMR modes, we observed a broad and relatively intense mode labeled P with the linear frequency-field dependence [Figs.~\ref{FIG:six}(a) and \ref{FIG:six}(c)]. This mode appears at frequencies, somewhat higher than the ones of the AFMR modes.
The coexistence of AFMR excitations and the mode P in the high-field range can be associated with spinon deconfinement observed in some quasi-one-dimensional quantum antiferromagnets~\cite{Lake_NatMat_2005_ChainScaling,ColdeaTennant_PRB_2003_Cs2CuCl4continua,SmirnovPovarov_PRB_2012_ESRordered,NawaHirai_PRR_2020_CROCspinons}.
This scenario is possible in COHB, since moderate magnetic fields significantly suppress magnetic order and, thus, enhance the quantum effects [see phase diagrams in Figs.~\ref{FIG:VSM}(d)-\ref{FIG:VSM}(f)].

The M and F modes show evidence of intricate interactions, depending on the direction of the magnetic field. In magnetic fields of about $13$ - $14$~T applied along the $a$ axis signatures of the avoided crossing appear. For $H\parallel b$, above about $12$~T, there is evidence for spin-wave damping upon interaction with the multiplet. At high magnetic fields, we also observe a number of weak resonances, labeled with crosses in Fig.~\ref{FIG:six}. Their origin remains an open question.

The correlation of the M modes parameters to the magnetic ordering is corroborated by the temperature dependence of the ESR signal. Figure~\ref{FIG:Tdep}(a) shows the temperature and field dependence of the ESR signal corresponding to mode M$_1$ ($H\parallel b$ at $133$~GHz), and Fig.~\ref{FIG:Tdep}(b) of the ESR signal corresponding to mode M$_2$ ($H\parallel c^\ast$ at $304.9$~GHz). Both modes demonstrate a rapid crossover to the paramagnetic resonance (with $g\sim 2.26$ and $2.03$, correspondingly) upon warming. For mode M$_1$ this can be examined in more detail, since its frequency-field dependence follows a law (at least up to $5$~T), typical for AFMR~\cite{GurevichMelkov_1996_Wavesbook}:

\begin{equation}
\nu_{1}=\sqrt{\Delta_1^{2}+(g\mu_0\mub H_\mathrm{res}/h)^2},\label{EQ:AFMR}
\end{equation}
where $h$ is the Planck constant. The temperature dependence of ESR field and linewidth at $133$~GHz, and the corresponding gap value $\Delta_{1}(T)$ are shown in Fig.~\ref{FIG:TdepM1}. The resonance field decreases rapidly upon cooling below $T_\mathrm{N}$, while the line becomes significantly narrower in the ordered phase. In contrast, at high temperatures, and especially near $T_{N}$ the line is very broad.

On the other hand, the F modes, resulting from bound states of spinons and magnons~\cite{ZhangZhao_PRL_2020_COHBspectrum}, show a largely different temperature dependence, as shown in Fig.~\ref{FIG:Tdep}(c) for $H\parallel b$ at $790$~GHz. Without any significant change in the respective position, the F$_\mathrm{B}$ resonances just loose the intensity and disappear. The same behavior was observed for the weaker satellite lines. All these lines become invisible at temperatures, corresponding to the energy scale of interchain coupling $J_{3}/k_B\sim T_\mathrm{N}$.

\section{Summary}

In summary, we performed systematic high-field magnetization and ESR studies of the alternating FM-AFM spin-$1/2$ chain compound \COHB. We observed a magnetization plateau with the magnetic moment close to one-half of the saturation value. This suggests the full polarization of the FM chain sublattice, while the large AFM exchange coupling keeps the AFM chain sublattice only weakly magnetized. We measured the angular dependence of the spin-reorientation transition, evidencing that the spin reorientation occurs within the FM chain sublattice.

Our ESR experiments indicate a very rich excitation spectrum in COHB. In addition to the high-energy multiplet (originating from magnon-spinon interactions) we observed two modes of antiferromagnetic resonance, a consequence of the long-range magnetic ordering below $T_\mathrm{N}=9.3$~K.

On the other hand, we observed a collapse of the magnetically ordered phase in applied magnetic field. We measured the $H-T$ phase diagram for three principal magnetic field directions, revealing the anisotropic nature of the order-disorder phase boundaries. These findings urge for careful accounting of possible small-term components of the COHB spin Hamiltonian, including the Dzyaloshinskii--Moriya components of the interchain couplings.

\acknowledgments

This work was supported by the Deutsche Forschungsgemeinschaft through the W\"{u}rzburg-Dresden Cluster of Excellence on Complexity and Topology in Quantum Matter - $ct.qmat$ (EXC 2147, project No. 390858490) and the SFB 1143, as well as by HLD at HZDR, member of the European Magnetic Field Laboratory (EMFL). A portion of this work was performed at the National High Magnetic Field Laboratory, which is supported by the National Science Foundation Cooperative Agreement No.~DMR-1644779 and the State of Florida. Z.Y.Z. is grateful for the support from the National Natural Science Foundation of China (Grant No.~52072368). We thank M. Uhlarz (HZDR) for assistance with the PPMS system.

\emph{Note Added.} Recently we became aware of the work of Reinold~\emph{et al}.~\cite{ReinoldBerger_arXiv_2024_COHBHiib}, who studied the phase diagram of COHB for magnetic fields applied along the chain direction and observed a field-induced collapse of the magnetic order with $T_\mathrm{N}=9.3$~K. This finding is in line with our results.

\appendix

\begin{figure*}
\center
\includegraphics[width=0.8\textwidth]{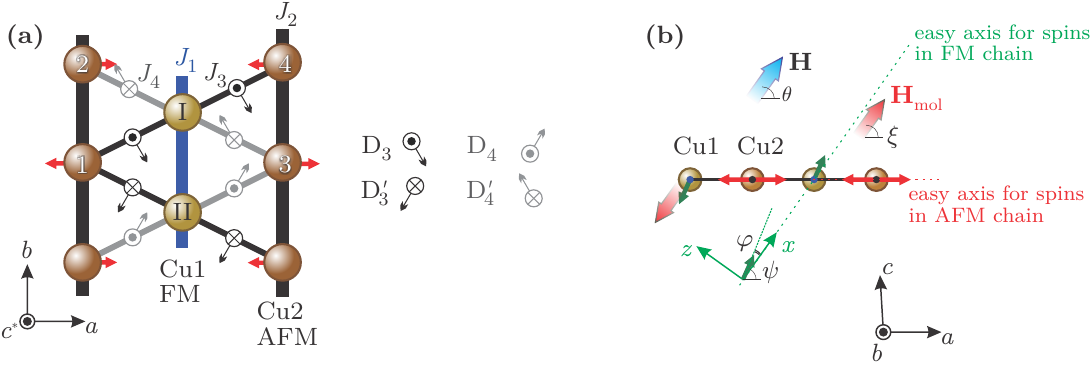}
\caption{(a) Mean-field theory scheme for the Cu$1$ sites in the FM chain subsystem. The red arrows show the directions of the spins in the AFM chains. Blue and black bars represent ferromagnetic $J_1$ and antiferromagnetic $J_2$ couplings, respectively. Interchain interactions $J_3$ and $J_4$ are denoted by the black and gray line segments; arrows are the DM vectors. (b) Details of the mean-field model geometry in $ac^{\ast}$ projection (see the text for angle definitions).}\label{FIG:MF}
\end{figure*}

\section{Interchain Dzyaloshinskii--Moriya interactions in COHB}\label{APP:DM}

In Fig.~\ref{FIG:MF}(a) we present the pattern of the possible interchain DM interactions in COHB (the symmetry analysis was performed with the help of \texttt{SpinW} package~\cite{TothLake_JPCM_2015_SpinW}). The bonds $J_3$ and $J_4$ host the DM interactions $\mathbf{D}_{3}$ and $\mathbf{D}_{4}$ correspondingly. The symmetry of COHB allows all components of the DM vector on these bonds. Both, $\mathbf{D}_{3}$ and $\mathbf{D}_{4}$, exist in two variants, differing by the sign of the $a$ and $c$ components (we always assume the bonds being ``directed'' from the Cu$1$ sites towards the Cu$2$ sites):
\begin{align}\nonumber
\mathbf{D}_{3}&=(D_{3a},-D_{3b},D_{3c}),\\ \nonumber
\mathbf{D}'_{3}&=(-D_{3a},-D_{3b},-D_{3c}),\\ \label{EQ:DMs}
\mathbf{D}_{4}&=(D_{4a},D_{4b},D_{4c}),\\ \nonumber
\mathbf{D}'_{4}&=(-D_{4a},D_{4b},-D_{4c}).\\ \nonumber
\end{align}

\section{Mean-field model for spin reorientation in COHB}\label{APP:SF}

We consider the influence of the AFM chain sublattice on the FM chains. Let us start with writing down the interactions acting on the Cu$1$ site I in the FM chain, as illustrated in Fig.~\ref{FIG:MF}(a). The corresponding spin Hamiltonian is:

\begin{align}\nonumber
&\hamilt_\mathrm{I}=\hamilt_\mathrm{intrachain}
+J_{3}\spop_\mathrm{I}\cdot(\spop_{1}+\spop_{4})
+J_{4}\spop_\mathrm{I}\cdot(\spop_{2}+\spop_{3})\\
&-\mathbf{D}_{3}\cdot[\spop_\mathrm{I}\times (\spop_{1}+\spop_{4})]
-\mathbf{D}'_{4}\cdot[\spop_\mathrm{I}\times (\spop_{2}+\spop_{3})].
\label{EQ:ToySF1}
\end{align}

We replace the spin operators related to the AFM Cu$2$ ions with the corresponding expectation values:
$$\aver{\spop_{1}}=\aver{\spop_{4}}=\begin{pmatrix}-S\\0\\0\end{pmatrix},
\aver{\spop_{2}}=\aver{\spop_{3}}=\begin{pmatrix}S\\0\\0\end{pmatrix}.$$
The combination of the strong $J_2$ exchange and its easy-axis anisotropy ($a$ as preferred direction) completely fixes the staggered arrangement of the Cu$2$ spins. Possible effects of the magnetic field on these spins are neglected, as we are considering the small-field limit only.
The resulting version of the Hamiltonian~(\ref{EQ:ToySF1}) reads as:

\begin{align}\nonumber
\hamilt_\mathrm{I}=\hamilt_\mathrm{intrachain}-\spop_\mathrm{I}\cdot\begin{pmatrix}2S\\0\\0\end{pmatrix}(J_{3}-J_{4})\\
+\spop_\mathrm{I}\cdot\begin{pmatrix}2S\\0\\0\end{pmatrix}\times(\mathbf{D}_{3}-\mathbf{D}'_{4}).
\label{EQ:ToySF2mf}
\end{align}
The last two terms are equivalent to the effective molecular field:
\begin{equation}\label{EQ:fieldFM2}
g\mub\mu_{0}\mathbf{H}^\mathrm{mol}_\mathrm{I}=2S\begin{pmatrix}J_{3}-J_{4}\\-D_{3c}-D_{4c}\\-D_{3b}-D_{4b}\end{pmatrix}.
\end{equation}
This is the effective field experienced by Cu$1$ on site I of the FM chain. The molecular field at site II is obtained by replacing $\mathbf{D}_{3}\rightarrow\mathbf{D}'_{3}$ and $\mathbf{D}'_{4}\rightarrow\mathbf{D}_{4}$ [as follows from the bond scheme in Fig.~\ref{FIG:MF}(a)]:

\begin{equation}\label{EQ:fieldFM1}
g\mub\mu_{0}\mathbf{H}^\mathrm{mol}_\mathrm{II}=2S\begin{pmatrix}J_{3}-J_{4}\\D_{3c}+D_{4c}\\-D_{4b}-D_{3b}\end{pmatrix}
\end{equation}
Thus, the effective molecular field experienced by the Cu$1$ spins in the FM chains consists of a uniform component in the $ac$ plane and staggered component along $b$.
Between neighboring FM chains these fields would be given by replacing $S\rightarrow -S$. Since there is no reports of a staggered spin component within the FM chains, we will assume the $D_{c}$ components to be negligible. This leaves only the uniform part of $\mathbf{H}^\mathrm{mol}$, with the sign alternating between neighboring FM chains. This leads to the Cu1 spin arrangement as proposed for the zero-field structure~\cite{ZhangZhao_PRL_2020_COHBspectrum}.

The intrachain Hamiltonian from Eq.~(\ref{EQ:ToySF1}) can be written as:

\begin{equation}
\hamilt_\mathrm{intrachain}=\sum\limits_{\alpha,\beta=x,y,z}2J_{1}^{\alpha\beta}\spin_\mathrm{I}^{\alpha}\cdot\aver{\spin^\beta_\mathrm{II}}.
\end{equation}
Here, $\aver{\spop_\mathrm{II}}$ is the spin expectation value for site I's neighbors within the chain. Thus, the full Hamiltonian of a single Cu$1$ spin is:

\begin{equation}
\hamilt_\mathrm{I}=\sum\limits_{\alpha,\beta}2J_{1}^{\alpha\beta}\spin^{\alpha}_\mathrm{I}\cdot\aver{\spin^\beta_\mathrm{II}}-g\mub\mu_{0}(\mathbf{H}\pm\mathbf{H}_\mathrm{mol})\spop_\mathrm{I}.\label{EQ:moreFinalHamilt}
\end{equation}
Here, the $\pm$ sign of the interchain molecular field refers to neighboring chains.

Let us assume that the local anisotropy axis of the FM chain lies within the $ac$ plane, at an angle $\psi$ with respect to the $a$ direction of the crystal. We select this direction as $x$, $b$ as $y$, and complement it with the orthogonal direction $z$. In this reference frame, the $J_{1}$ tensor is
\begin{equation}
J^{\alpha\beta}_\mathrm{1}=J_\mathrm{1}\begin{pmatrix}
1+\delta & 0 & 0\\
0 & 1 & 0 \\
0 & 0& 1
\end{pmatrix},
\end{equation}
and the effective interchain field is
\begin{equation}
\mathbf{H}_\mathrm{mol}=H_\mathrm{mol}\begin{pmatrix}
\cos{(\psi-\xi)}\\ 0 \\ -\sin{(\psi-\xi)}
\end{pmatrix}.
\end{equation}
Here, $\tan{\xi}=(-D_{3b}-D_{4b})/(J_{3}-J_{4})$ is the effective field angle, resulting from the interplay of the DM interactions and symmetric exchange, and $H_\mathrm{mol}=2S(g\mub\mu_0)^{-1}\sqrt{(J_{3}-J_{4})^2+(D_{3b}+D_{4b})^2}$ is the corresponding molecular field magnitude.
Let the external field be aligned at an angle $\theta$ to the $a$-axis in the $ac$ plane, written as:

\begin{equation}
\mathbf{H}=H\begin{pmatrix}
\cos{(\psi-\theta)}\\ 0 \\ -\sin{(\psi-\theta)}
\end{pmatrix}
\end{equation}
in the $xyz$ reference frame. Finally, the spin operators in the semiclassical approximation are

\begin{equation}
\spop_\mathrm{I,II}=\aver{\spop_\mathrm{I,II}}=S\begin{pmatrix}
\cos{\varphi}\\ 0 \\ \sin{\varphi}
\end{pmatrix},
\end{equation}
with the angle $\varphi$ characterizing the possible tilt away from the anisotropy axis $x$. Thus, the Hamiltonian~(\ref{EQ:moreFinalHamilt}) corresponds to the energy:

\begin{widetext}
\begin{equation}
E_{\pm}(\varphi)=-2J_{1}S^2-2J_{1}S^{2}\delta\cos^{2}\varphi\pm g\mub\mu_0 H_\mathrm{mol}S\cos{(\varphi+\psi-\xi)}-g\mub\mu_0 HS \cos(\psi+\varphi-\theta),\label{EQ:MFSFmodel0}
\end{equation}
\end{widetext}
with different $H_\mathrm{mol}$ sign for neighboring chains.

We consider independent $\varphi$ angles for neighboring chains, $\varphi_{+}$ and $\varphi_{-}$, respectively. Then, the full energy we need to minimize is:

\begin{equation}
E(\varphi_{+},\varphi_{-})=E_{+}(\varphi_{+})+E_{-}(\varphi_{-}).\label{EQ:MFSFmodel}
\end{equation}
This energy can be minimized for a given external field and a set of intrinsic parameters $2J_\mathrm{1}\delta$, $H_\mathrm{mol}$, $\psi$, and $\xi$. Thus, we can determine the optimal spin configuration in the FM chains and obtain the magnetization curve for the Cu$1$ chain sublattice.
In Fig.~5(c) of the main text we show the experimentally measured $H_\mathrm{SR}$ field as function of orientation, overlayed with the spin susceptibility in the Cu$1$ chain sublattice, which is $-d^2 E/ dH^2$ derivative of the energy Eq.~(\ref{EQ:MFSFmodel}) at optimal values of angles $\varphi_{+}$, $\varphi_{-}$.

For a good agreement with the COHB data, both angles $\psi$ and $\xi$ need to be close to $45^\circ$.
The minimum in $H_\mathrm{SR}(\theta)$ is located at $\theta\simeq\psi$. Thus, the experimentally found minimum at $45^\circ$ determines the orientation of the intrachain anisotropy axis. The angle $\xi$ (and, hence, strength of the DM interaction) has a more subtle effect. A molecular field pointing away from the intrachain anisotropy axis direction keeps the critical-field minimum near $\theta=\psi$, but makes the $H_\mathrm{SR}(\theta)$ dependence asymmetric with respect to this angle. Hence, only $\xi\simeq 45^\circ$ (and, consequently, nearly equal contributions from Heisenberg and DM interactions) allows a good agreement with the nearly-symmetric angular dependence obtained experimentally.

We find that experimental $H_\mathrm{SR}(\theta)$ is best fit with the effective parameters $2S^2|J_{1}|\delta/\kb\simeq3.2$~K and $g\mub\mu_{0}H_\mathrm{mol}/\kb\simeq3.5$~K, with both $\xi\simeq\psi\simeq45^{\circ}$. This is consistent with $J_\mathrm{1}/\kb=-30$~K and $\delta=0.17$ from linear spin-wave theory~\cite{ZhangZhao_PRL_2020_COHBspectrum}. The difference $(J_{3}-J_{4})/\kb$ is roughly the same as sum $|D_{3b}+D_{4b}|/\kb$, both of them being about $2.5$~K. The sign of $D_{3b}+D_{4b}$ should be negative; then the direction of the molecular field would fully align with the direction of the anisotropy axis at $J_{1}$ bond. Reversely, the staggered molecular field experienced by the spins in the AFM chain would be strictly along $a$ in that case.

\bibliography{ESR_v20.bib}

\begin{thebibliography}{36}%
\makeatletter
\providecommand \@ifxundefined [1]{%
 \@ifx{#1\undefined}
}%
\providecommand \@ifnum [1]{%
 \ifnum #1\expandafter \@firstoftwo
 \else \expandafter \@secondoftwo
 \fi
}%
\providecommand \@ifx [1]{%
 \ifx #1\expandafter \@firstoftwo
 \else \expandafter \@secondoftwo
 \fi
}%
\providecommand \natexlab [1]{#1}%
\providecommand \enquote  [1]{``#1''}%
\providecommand \bibnamefont  [1]{#1}%
\providecommand \bibfnamefont [1]{#1}%
\providecommand \citenamefont [1]{#1}%
\providecommand \href@noop [0]{\@secondoftwo}%
\providecommand \href [0]{\begingroup \@sanitize@url \@href}%
\providecommand \@href[1]{\@@startlink{#1}\@@href}%
\providecommand \@@href[1]{\endgroup#1\@@endlink}%
\providecommand \@sanitize@url [0]{\catcode `\\12\catcode `\$12\catcode
  `\&12\catcode `\#12\catcode `\^12\catcode `\_12\catcode `\%12\relax}%
\providecommand \@@startlink[1]{}%
\providecommand \@@endlink[0]{}%
\providecommand \url  [0]{\begingroup\@sanitize@url \@url }%
\providecommand \@url [1]{\endgroup\@href {#1}{\urlprefix }}%
\providecommand \urlprefix  [0]{URL }%
\providecommand \Eprint [0]{\href }%
\providecommand \doibase [0]{https://doi.org/}%
\providecommand \selectlanguage [0]{\@gobble}%
\providecommand \bibinfo  [0]{\@secondoftwo}%
\providecommand \bibfield  [0]{\@secondoftwo}%
\providecommand \translation [1]{[#1]}%
\providecommand \BibitemOpen [0]{}%
\providecommand \bibitemStop [0]{}%
\providecommand \bibitemNoStop [0]{.\EOS\space}%
\providecommand \EOS [0]{\spacefactor3000\relax}%
\providecommand \BibitemShut  [1]{\csname bibitem#1\endcsname}%
\let\auto@bib@innerbib\@empty
\bibitem [{\citenamefont {Inosov}(2018)}]{Inosov_AdvPhys_2018_MineralReview}%
  \BibitemOpen
  \bibfield  {author} {\bibinfo {author} {\bibfnamefont {D.}~\bibnamefont
  {Inosov}},\ }\bibfield  {title} {\bibinfo {title} {Quantum magnetism in
  minerals},\ }\href {https://doi.org/10.1080/00018732.2018.1571986} {\bibfield
   {journal} {\bibinfo  {journal} {Adv. Phys.}\ }\textbf {\bibinfo {volume}
  {67}},\ \bibinfo {pages} {149} (\bibinfo {year} {2018})}\BibitemShut
  {NoStop}%
\bibitem [{\citenamefont {Willenberg}\ \emph {et~al.}(2012)\citenamefont
  {Willenberg}, \citenamefont {Sch\"apers}, \citenamefont {Rule}, \citenamefont
  {S\"ullow}, \citenamefont {Reehuis}, \citenamefont {Ryll}, \citenamefont
  {Klemke}, \citenamefont {Kiefer}, \citenamefont {Schottenhamel},
  \citenamefont {B\"uchner}, \citenamefont {Ouladdiaf}, \citenamefont {Uhlarz},
  \citenamefont {Beyer}, \citenamefont {Wosnitza},\ and\ \citenamefont
  {Wolter}}]{Willenberg_PRL_2012_LinariteFrustrated}%
  \BibitemOpen
  \bibfield  {author} {\bibinfo {author} {\bibfnamefont {B.}~\bibnamefont
  {Willenberg}}, \bibinfo {author} {\bibfnamefont {M.}~\bibnamefont
  {Sch\"apers}}, \bibinfo {author} {\bibfnamefont {K.~C.}\ \bibnamefont
  {Rule}}, \bibinfo {author} {\bibfnamefont {S.}~\bibnamefont {S\"ullow}},
  \bibinfo {author} {\bibfnamefont {M.}~\bibnamefont {Reehuis}}, \bibinfo
  {author} {\bibfnamefont {H.}~\bibnamefont {Ryll}}, \bibinfo {author}
  {\bibfnamefont {B.}~\bibnamefont {Klemke}}, \bibinfo {author} {\bibfnamefont
  {K.}~\bibnamefont {Kiefer}}, \bibinfo {author} {\bibfnamefont
  {W.}~\bibnamefont {Schottenhamel}}, \bibinfo {author} {\bibfnamefont
  {B.}~\bibnamefont {B\"uchner}}, \bibinfo {author} {\bibfnamefont
  {B.}~\bibnamefont {Ouladdiaf}}, \bibinfo {author} {\bibfnamefont
  {M.}~\bibnamefont {Uhlarz}}, \bibinfo {author} {\bibfnamefont
  {R.}~\bibnamefont {Beyer}}, \bibinfo {author} {\bibfnamefont
  {J.}~\bibnamefont {Wosnitza}},\ and\ \bibinfo {author} {\bibfnamefont
  {A.~U.~B.}\ \bibnamefont {Wolter}},\ }\bibfield  {title} {\bibinfo {title}
  {{Magnetic Frustration in a Quantum Spin Chain: The Case of Linarite
  ${\mathrm{PbCuSO}}_{4}(\mathrm{OH}{)}_{2}$}},\ }\href
  {https://doi.org/10.1103/PhysRevLett.108.117202} {\bibfield  {journal}
  {\bibinfo  {journal} {Phys. Rev. Lett.}\ }\textbf {\bibinfo {volume} {108}},\
  \bibinfo {pages} {117202} (\bibinfo {year} {2012})}\BibitemShut {NoStop}%
\bibitem [{\citenamefont {Willenberg}\ \emph {et~al.}(2016)\citenamefont
  {Willenberg}, \citenamefont {Sch\"apers}, \citenamefont {Wolter},
  \citenamefont {Drechsler}, \citenamefont {Reehuis}, \citenamefont {Hoffmann},
  \citenamefont {B\"uchner}, \citenamefont {Studer}, \citenamefont {Rule},
  \citenamefont {Ouladdiaf}, \citenamefont {S\"ullow},\ and\ \citenamefont
  {Nishimoto}}]{Willenberg_PRL_2016_LinariteSDWs}%
  \BibitemOpen
  \bibfield  {author} {\bibinfo {author} {\bibfnamefont {B.}~\bibnamefont
  {Willenberg}}, \bibinfo {author} {\bibfnamefont {M.}~\bibnamefont
  {Sch\"apers}}, \bibinfo {author} {\bibfnamefont {A.~U.~B.}\ \bibnamefont
  {Wolter}}, \bibinfo {author} {\bibfnamefont {S.-L.}\ \bibnamefont
  {Drechsler}}, \bibinfo {author} {\bibfnamefont {M.}~\bibnamefont {Reehuis}},
  \bibinfo {author} {\bibfnamefont {J.-U.}\ \bibnamefont {Hoffmann}}, \bibinfo
  {author} {\bibfnamefont {B.}~\bibnamefont {B\"uchner}}, \bibinfo {author}
  {\bibfnamefont {A.~J.}\ \bibnamefont {Studer}}, \bibinfo {author}
  {\bibfnamefont {K.~C.}\ \bibnamefont {Rule}}, \bibinfo {author}
  {\bibfnamefont {B.}~\bibnamefont {Ouladdiaf}}, \bibinfo {author}
  {\bibfnamefont {S.}~\bibnamefont {S\"ullow}},\ and\ \bibinfo {author}
  {\bibfnamefont {S.}~\bibnamefont {Nishimoto}},\ }\bibfield  {title} {\bibinfo
  {title} {{Complex Field-Induced States in Linarite
  ${\mathrm{PbCuSO}}_{4}(\mathrm{OH}{)}_{2}$ with a Variety of High-Order
  Exotic Spin-Density Wave States}},\ }\href
  {https://doi.org/10.1103/PhysRevLett.116.047202} {\bibfield  {journal}
  {\bibinfo  {journal} {Phys. Rev. Lett.}\ }\textbf {\bibinfo {volume} {116}},\
  \bibinfo {pages} {047202} (\bibinfo {year} {2016})}\BibitemShut {NoStop}%
\bibitem [{\citenamefont {Povarov}\ \emph {et~al.}(2016)\citenamefont
  {Povarov}, \citenamefont {Feng},\ and\ \citenamefont
  {Zheludev}}]{PovarovFeng_PRB_2016_LinariteMF}%
  \BibitemOpen
  \bibfield  {author} {\bibinfo {author} {\bibfnamefont {{\relax K.
  Yu}.}~\bibnamefont {Povarov}}, \bibinfo {author} {\bibfnamefont
  {Y.}~\bibnamefont {Feng}},\ and\ \bibinfo {author} {\bibfnamefont
  {A.}~\bibnamefont {Zheludev}},\ }\bibfield  {title} {\bibinfo {title}
  {Multiferroic phases of the frustrated quantum spin-chain compound
  linarite},\ }\href {https://doi.org/10.1103/PhysRevB.94.214409} {\bibfield
  {journal} {\bibinfo  {journal} {Phys. Rev. B}\ }\textbf {\bibinfo {volume}
  {94}},\ \bibinfo {pages} {214409} (\bibinfo {year} {2016})}\BibitemShut
  {NoStop}%
\bibitem [{\citenamefont {Cemal}\ \emph {et~al.}(2018)\citenamefont {Cemal},
  \citenamefont {Enderle}, \citenamefont {Kremer}, \citenamefont {F\aa{}k},
  \citenamefont {Ressouche}, \citenamefont {Goff}, \citenamefont {Gvozdikova},
  \citenamefont {Zhitomirsky},\ and\ \citenamefont
  {Ziman}}]{CemalEnderle_PRL_2018_Linarite}%
  \BibitemOpen
  \bibfield  {author} {\bibinfo {author} {\bibfnamefont {E.}~\bibnamefont
  {Cemal}}, \bibinfo {author} {\bibfnamefont {M.}~\bibnamefont {Enderle}},
  \bibinfo {author} {\bibfnamefont {R.~K.}\ \bibnamefont {Kremer}}, \bibinfo
  {author} {\bibfnamefont {B.}~\bibnamefont {F\aa{}k}}, \bibinfo {author}
  {\bibfnamefont {E.}~\bibnamefont {Ressouche}}, \bibinfo {author}
  {\bibfnamefont {J.~P.}\ \bibnamefont {Goff}}, \bibinfo {author}
  {\bibfnamefont {M.~V.}\ \bibnamefont {Gvozdikova}}, \bibinfo {author}
  {\bibfnamefont {M.~E.}\ \bibnamefont {Zhitomirsky}},\ and\ \bibinfo {author}
  {\bibfnamefont {T.}~\bibnamefont {Ziman}},\ }\bibfield  {title} {\bibinfo
  {title} {{Field-induced States and Excitations in the Quasicritical
  Spin-$1/2$ Chain Linarite}},\ }\href
  {https://doi.org/10.1103/PhysRevLett.120.067203} {\bibfield  {journal}
  {\bibinfo  {journal} {Phys. Rev. Lett.}\ }\textbf {\bibinfo {volume} {120}},\
  \bibinfo {pages} {067203} (\bibinfo {year} {2018})}\BibitemShut {NoStop}%
\bibitem [{\citenamefont {Mendels}\ \emph {et~al.}(2007)\citenamefont
  {Mendels}, \citenamefont {Bert}, \citenamefont {de~Vries}, \citenamefont
  {Olariu}, \citenamefont {Harrison}, \citenamefont {Duc}, \citenamefont
  {Trombe}, \citenamefont {Lord}, \citenamefont {Amato},\ and\ \citenamefont
  {Baines}}]{MendelsBert_PRL_2007_HerbertNoLRO}%
  \BibitemOpen
  \bibfield  {author} {\bibinfo {author} {\bibfnamefont {P.}~\bibnamefont
  {Mendels}}, \bibinfo {author} {\bibfnamefont {F.}~\bibnamefont {Bert}},
  \bibinfo {author} {\bibfnamefont {M.~A.}\ \bibnamefont {de~Vries}}, \bibinfo
  {author} {\bibfnamefont {A.}~\bibnamefont {Olariu}}, \bibinfo {author}
  {\bibfnamefont {A.}~\bibnamefont {Harrison}}, \bibinfo {author}
  {\bibfnamefont {F.}~\bibnamefont {Duc}}, \bibinfo {author} {\bibfnamefont
  {J.~C.}\ \bibnamefont {Trombe}}, \bibinfo {author} {\bibfnamefont {J.~S.}\
  \bibnamefont {Lord}}, \bibinfo {author} {\bibfnamefont {A.}~\bibnamefont
  {Amato}},\ and\ \bibinfo {author} {\bibfnamefont {C.}~\bibnamefont
  {Baines}},\ }\bibfield  {title} {\bibinfo {title} {{Quantum Magnetism in the
  Paratacamite Family: Towards an Ideal Kagom\'e Lattice}},\ }\href
  {https://doi.org/10.1103/PhysRevLett.98.077204} {\bibfield  {journal}
  {\bibinfo  {journal} {Phys. Rev. Lett.}\ }\textbf {\bibinfo {volume} {98}},\
  \bibinfo {pages} {077204} (\bibinfo {year} {2007})}\BibitemShut {NoStop}%
\bibitem [{\citenamefont {Norman}(2016)}]{Norman_RMP_2016_HerbReview}%
  \BibitemOpen
  \bibfield  {author} {\bibinfo {author} {\bibfnamefont {M.~R.}\ \bibnamefont
  {Norman}},\ }\bibfield  {title} {\bibinfo {title} {{Colloquium:
  Herbertsmithite and the search for the quantum spin liquid}},\ }\href
  {https://doi.org/10.1103/RevModPhys.88.041002} {\bibfield  {journal}
  {\bibinfo  {journal} {Rev. Mod. Phys.}\ }\textbf {\bibinfo {volume} {88}},\
  \bibinfo {pages} {041002} (\bibinfo {year} {2016})}\BibitemShut {NoStop}%
\bibitem [{\citenamefont {Barth\'elemy}\ \emph {et~al.}(2022)\citenamefont
  {Barth\'elemy}, \citenamefont {Demuer}, \citenamefont {Marcenat},
  \citenamefont {Klein}, \citenamefont {Bernu}, \citenamefont {Messio},
  \citenamefont {Vel\'azquez}, \citenamefont {Kermarrec}, \citenamefont
  {Bert},\ and\ \citenamefont
  {Mendels}}]{BarthelemyDemuer_PRX_2022_HerbertsmCp}%
  \BibitemOpen
  \bibfield  {author} {\bibinfo {author} {\bibfnamefont {Q.}~\bibnamefont
  {Barth\'elemy}}, \bibinfo {author} {\bibfnamefont {A.}~\bibnamefont
  {Demuer}}, \bibinfo {author} {\bibfnamefont {C.}~\bibnamefont {Marcenat}},
  \bibinfo {author} {\bibfnamefont {T.}~\bibnamefont {Klein}}, \bibinfo
  {author} {\bibfnamefont {B.}~\bibnamefont {Bernu}}, \bibinfo {author}
  {\bibfnamefont {L.}~\bibnamefont {Messio}}, \bibinfo {author} {\bibfnamefont
  {M.}~\bibnamefont {Vel\'azquez}}, \bibinfo {author} {\bibfnamefont
  {E.}~\bibnamefont {Kermarrec}}, \bibinfo {author} {\bibfnamefont
  {F.}~\bibnamefont {Bert}},\ and\ \bibinfo {author} {\bibfnamefont
  {P.}~\bibnamefont {Mendels}},\ }\bibfield  {title} {\bibinfo {title}
  {{Specific Heat of the Kagome Antiferromagnet Herbertsmithite in High
  Magnetic Fields}},\ }\href {https://doi.org/10.1103/PhysRevX.12.011014}
  {\bibfield  {journal} {\bibinfo  {journal} {Phys. Rev. X}\ }\textbf {\bibinfo
  {volume} {12}},\ \bibinfo {pages} {011014} (\bibinfo {year}
  {2022})}\BibitemShut {NoStop}%
\bibitem [{\citenamefont {Heinze}\ \emph {et~al.}(2021)\citenamefont {Heinze},
  \citenamefont {Jeschke}, \citenamefont {Mazin}, \citenamefont
  {Metavitsiadis}, \citenamefont {Reehuis}, \citenamefont {Feyerherm},
  \citenamefont {Hoffmann}, \citenamefont {Bartkowiak}, \citenamefont
  {Prokhnenko}, \citenamefont {Wolter}, \citenamefont {Ding}, \citenamefont
  {Zapf}, \citenamefont {Corval\'an~Moya}, \citenamefont {Weickert},
  \citenamefont {Jaime}, \citenamefont {Rule}, \citenamefont {Menzel},
  \citenamefont {Valent\'{\i}}, \citenamefont {Brenig},\ and\ \citenamefont
  {S\"ullow}}]{HeinzeJeshke_PRL_2021_AtacamitePlateau}%
  \BibitemOpen
  \bibfield  {author} {\bibinfo {author} {\bibfnamefont {L.}~\bibnamefont
  {Heinze}}, \bibinfo {author} {\bibfnamefont {H.~O.}\ \bibnamefont {Jeschke}},
  \bibinfo {author} {\bibfnamefont {I.~I.}\ \bibnamefont {Mazin}}, \bibinfo
  {author} {\bibfnamefont {A.}~\bibnamefont {Metavitsiadis}}, \bibinfo {author}
  {\bibfnamefont {M.}~\bibnamefont {Reehuis}}, \bibinfo {author} {\bibfnamefont
  {R.}~\bibnamefont {Feyerherm}}, \bibinfo {author} {\bibfnamefont {J.-U.}\
  \bibnamefont {Hoffmann}}, \bibinfo {author} {\bibfnamefont {M.}~\bibnamefont
  {Bartkowiak}}, \bibinfo {author} {\bibfnamefont {O.}~\bibnamefont
  {Prokhnenko}}, \bibinfo {author} {\bibfnamefont {A.~U.~B.}\ \bibnamefont
  {Wolter}}, \bibinfo {author} {\bibfnamefont {X.}~\bibnamefont {Ding}},
  \bibinfo {author} {\bibfnamefont {V.~S.}\ \bibnamefont {Zapf}}, \bibinfo
  {author} {\bibfnamefont {C.}~\bibnamefont {Corval\'an~Moya}}, \bibinfo
  {author} {\bibfnamefont {F.}~\bibnamefont {Weickert}}, \bibinfo {author}
  {\bibfnamefont {M.}~\bibnamefont {Jaime}}, \bibinfo {author} {\bibfnamefont
  {K.~C.}\ \bibnamefont {Rule}}, \bibinfo {author} {\bibfnamefont
  {D.}~\bibnamefont {Menzel}}, \bibinfo {author} {\bibfnamefont
  {R.}~\bibnamefont {Valent\'{\i}}}, \bibinfo {author} {\bibfnamefont
  {W.}~\bibnamefont {Brenig}},\ and\ \bibinfo {author} {\bibfnamefont
  {S.}~\bibnamefont {S\"ullow}},\ }\bibfield  {title} {\bibinfo {title}
  {{Magnetization Process of Atacamite: A Case of Weakly Coupled $S=1/2$
  Sawtooth Chains}},\ }\href {https://doi.org/10.1103/PhysRevLett.126.207201}
  {\bibfield  {journal} {\bibinfo  {journal} {Phys. Rev. Lett.}\ }\textbf
  {\bibinfo {volume} {126}},\ \bibinfo {pages} {207201} (\bibinfo {year}
  {2021})}\BibitemShut {NoStop}%
\bibitem [{\citenamefont {Kulbakov}\ \emph
  {et~al.}(2022{\natexlab{a}})\citenamefont {Kulbakov}, \citenamefont
  {Kononenko}, \citenamefont {Nishimoto}, \citenamefont {Stahl}, \citenamefont
  {Chakkingal}, \citenamefont {Feig}, \citenamefont {Gumeniuk}, \citenamefont
  {Skourski}, \citenamefont {Bhaskaran}, \citenamefont {Zvyagin}, \citenamefont
  {Embs}, \citenamefont {Puente-Orench}, \citenamefont {Wildes}, \citenamefont
  {Geck}, \citenamefont {Janson}, \citenamefont {Inosov},\ and\ \citenamefont
  {Peets}}]{KulbakovKononenko_PRB_2022_Antlerite}%
  \BibitemOpen
  \bibfield  {author} {\bibinfo {author} {\bibfnamefont {A.~A.}\ \bibnamefont
  {Kulbakov}}, \bibinfo {author} {\bibfnamefont {D.~Y.}\ \bibnamefont
  {Kononenko}}, \bibinfo {author} {\bibfnamefont {S.}~\bibnamefont
  {Nishimoto}}, \bibinfo {author} {\bibfnamefont {Q.}~\bibnamefont {Stahl}},
  \bibinfo {author} {\bibfnamefont {A.~M.}\ \bibnamefont {Chakkingal}},
  \bibinfo {author} {\bibfnamefont {M.}~\bibnamefont {Feig}}, \bibinfo {author}
  {\bibfnamefont {R.}~\bibnamefont {Gumeniuk}}, \bibinfo {author}
  {\bibfnamefont {Y.}~\bibnamefont {Skourski}}, \bibinfo {author}
  {\bibfnamefont {L.}~\bibnamefont {Bhaskaran}}, \bibinfo {author}
  {\bibfnamefont {S.~A.}\ \bibnamefont {Zvyagin}}, \bibinfo {author}
  {\bibfnamefont {J.~P.}\ \bibnamefont {Embs}}, \bibinfo {author}
  {\bibfnamefont {I.}~\bibnamefont {Puente-Orench}}, \bibinfo {author}
  {\bibfnamefont {A.}~\bibnamefont {Wildes}}, \bibinfo {author} {\bibfnamefont
  {J.}~\bibnamefont {Geck}}, \bibinfo {author} {\bibfnamefont {O.}~\bibnamefont
  {Janson}}, \bibinfo {author} {\bibfnamefont {D.~S.}\ \bibnamefont {Inosov}},\
  and\ \bibinfo {author} {\bibfnamefont {D.~C.}\ \bibnamefont {Peets}},\
  }\bibfield  {title} {\bibinfo {title} {{Coupled frustrated ferromagnetic and
  antiferromagnetic quantum spin chains in the quasi-one-dimensional mineral
  antlerite ${\mathrm{Cu}}_{3}{\mathrm{SO}}_{4}$(OH)${}_{4}$}},\ }\href
  {https://doi.org/10.1103/PhysRevB.106.L020405} {\bibfield  {journal}
  {\bibinfo  {journal} {Phys. Rev. B}\ }\textbf {\bibinfo {volume} {106}},\
  \bibinfo {pages} {L020405} (\bibinfo {year}
  {2022}{\natexlab{a}})}\BibitemShut {NoStop}%
\bibitem [{\citenamefont {Kulbakov}\ \emph
  {et~al.}(2022{\natexlab{b}})\citenamefont {Kulbakov}, \citenamefont
  {Sadrollahi}, \citenamefont {Rasch}, \citenamefont {Avdeev}, \citenamefont
  {Ga\ss{}}, \citenamefont {Corredor~Bohorquez}, \citenamefont {Wolter},
  \citenamefont {Feig}, \citenamefont {Gumeniuk}, \citenamefont {Poddig},
  \citenamefont {St\"otzer}, \citenamefont {Litterst}, \citenamefont
  {Puente-Orench}, \citenamefont {Wildes}, \citenamefont {Weschke},
  \citenamefont {Geck}, \citenamefont {Inosov},\ and\ \citenamefont
  {Peets}}]{KulbakovSadrollahi_PRB_2022_AntleriteQ}%
  \BibitemOpen
  \bibfield  {author} {\bibinfo {author} {\bibfnamefont {A.~A.}\ \bibnamefont
  {Kulbakov}}, \bibinfo {author} {\bibfnamefont {E.}~\bibnamefont
  {Sadrollahi}}, \bibinfo {author} {\bibfnamefont {F.}~\bibnamefont {Rasch}},
  \bibinfo {author} {\bibfnamefont {M.}~\bibnamefont {Avdeev}}, \bibinfo
  {author} {\bibfnamefont {S.}~\bibnamefont {Ga\ss{}}}, \bibinfo {author}
  {\bibfnamefont {L.~T.}\ \bibnamefont {Corredor~Bohorquez}}, \bibinfo {author}
  {\bibfnamefont {A.~U.~B.}\ \bibnamefont {Wolter}}, \bibinfo {author}
  {\bibfnamefont {M.}~\bibnamefont {Feig}}, \bibinfo {author} {\bibfnamefont
  {R.}~\bibnamefont {Gumeniuk}}, \bibinfo {author} {\bibfnamefont
  {H.}~\bibnamefont {Poddig}}, \bibinfo {author} {\bibfnamefont
  {M.}~\bibnamefont {St\"otzer}}, \bibinfo {author} {\bibfnamefont {F.~J.}\
  \bibnamefont {Litterst}}, \bibinfo {author} {\bibfnamefont {I.}~\bibnamefont
  {Puente-Orench}}, \bibinfo {author} {\bibfnamefont {A.}~\bibnamefont
  {Wildes}}, \bibinfo {author} {\bibfnamefont {E.}~\bibnamefont {Weschke}},
  \bibinfo {author} {\bibfnamefont {J.}~\bibnamefont {Geck}}, \bibinfo {author}
  {\bibfnamefont {D.~S.}\ \bibnamefont {Inosov}},\ and\ \bibinfo {author}
  {\bibfnamefont {D.~C.}\ \bibnamefont {Peets}},\ }\bibfield  {title} {\bibinfo
  {title} {{Incommensurate and multiple-$q$ magnetic misfit order in the
  frustrated quantum spin ladder material antlerite
  ${\mathrm{Cu}}_{3}{\mathrm{SO}}_{4}{(\mathrm{OH})}_{4}$}},\ }\href
  {https://doi.org/10.1103/PhysRevB.106.174431} {\bibfield  {journal} {\bibinfo
   {journal} {Phys. Rev. B}\ }\textbf {\bibinfo {volume} {106}},\ \bibinfo
  {pages} {174431} (\bibinfo {year} {2022}{\natexlab{b}})}\BibitemShut
  {NoStop}%
\bibitem [{\citenamefont {Diep}(2013)}]{Diep_2013_FrustBook}%
  \BibitemOpen
  \bibfield  {author} {\bibinfo {author} {\bibfnamefont {H.~T.}\ \bibnamefont
  {Diep}},\ }\href {https://doi.org/10.1142/8676} {\emph {\bibinfo {title}
  {{Frustrated Spin Systems}}}}\ (\bibinfo  {publisher} {World Scientific
  Publishing Co, Singapore},\ \bibinfo {year} {2013})\BibitemShut {NoStop}%
\bibitem [{\citenamefont
  {Starykh}(2015)}]{Starykh_RepPrPhys_2015_TriangularReview}%
  \BibitemOpen
  \bibfield  {author} {\bibinfo {author} {\bibfnamefont {O.~A.}\ \bibnamefont
  {Starykh}},\ }\bibfield  {title} {\bibinfo {title} {{Unusual ordered phases
  of highly frustrated magnets: a review}},\ }\href
  {https://doi.org/10.1088/0034-4885/78/5/052502} {\bibfield  {journal}
  {\bibinfo  {journal} {Rep. Prog. Phys.}\ }\textbf {\bibinfo {volume} {78}},\
  \bibinfo {pages} {052502} (\bibinfo {year} {2015})}\BibitemShut {NoStop}%
\bibitem [{\citenamefont {Zhao}\ \emph {et~al.}(2019)\citenamefont {Zhao},
  \citenamefont {Che}, \citenamefont {Chen}, \citenamefont {Wang},
  \citenamefont {Sun},\ and\ \citenamefont {He}}]{ZhaoChe_JPCM_2019_COHB}%
  \BibitemOpen
  \bibfield  {author} {\bibinfo {author} {\bibfnamefont {Z.~Y.}\ \bibnamefont
  {Zhao}}, \bibinfo {author} {\bibfnamefont {H.~L.}\ \bibnamefont {Che}},
  \bibinfo {author} {\bibfnamefont {R.}~\bibnamefont {Chen}}, \bibinfo {author}
  {\bibfnamefont {J.~F.}\ \bibnamefont {Wang}}, \bibinfo {author}
  {\bibfnamefont {X.~F.}\ \bibnamefont {Sun}},\ and\ \bibinfo {author}
  {\bibfnamefont {Z.~Z.}\ \bibnamefont {He}},\ }\bibfield  {title} {\bibinfo
  {title} {{Magnetism study on a triangular lattice antiferromagnet
  ${\mathrm{Cu}}_{2}{(\mathrm{OH})}_{3}\mathrm{Br}$}},\ }\href
  {https://doi.org/10.1088/1361-648X/ab1623} {\bibfield  {journal} {\bibinfo
  {journal} {J. Phys.: Cond. Matter}\ }\textbf {\bibinfo {volume} {31}},\
  \bibinfo {pages} {275801} (\bibinfo {year} {2019})}\BibitemShut {NoStop}%
\bibitem [{\citenamefont {Zhang}\ \emph {et~al.}(2020)\citenamefont {Zhang},
  \citenamefont {Zhao}, \citenamefont {Gautreau}, \citenamefont {Raczkowski},
  \citenamefont {Saha}, \citenamefont {Garlea}, \citenamefont {Cao},
  \citenamefont {Hong}, \citenamefont {Jeschke}, \citenamefont {Mahanti},
  \citenamefont {Birol}, \citenamefont {Assaad},\ and\ \citenamefont
  {Ke}}]{ZhangZhao_PRL_2020_COHBspectrum}%
  \BibitemOpen
  \bibfield  {author} {\bibinfo {author} {\bibfnamefont {H.}~\bibnamefont
  {Zhang}}, \bibinfo {author} {\bibfnamefont {Z.}~\bibnamefont {Zhao}},
  \bibinfo {author} {\bibfnamefont {D.}~\bibnamefont {Gautreau}}, \bibinfo
  {author} {\bibfnamefont {M.}~\bibnamefont {Raczkowski}}, \bibinfo {author}
  {\bibfnamefont {A.}~\bibnamefont {Saha}}, \bibinfo {author} {\bibfnamefont
  {V.~O.}\ \bibnamefont {Garlea}}, \bibinfo {author} {\bibfnamefont
  {H.}~\bibnamefont {Cao}}, \bibinfo {author} {\bibfnamefont {T.}~\bibnamefont
  {Hong}}, \bibinfo {author} {\bibfnamefont {H.~O.}\ \bibnamefont {Jeschke}},
  \bibinfo {author} {\bibfnamefont {S.~D.}\ \bibnamefont {Mahanti}}, \bibinfo
  {author} {\bibfnamefont {T.}~\bibnamefont {Birol}}, \bibinfo {author}
  {\bibfnamefont {F.~F.}\ \bibnamefont {Assaad}},\ and\ \bibinfo {author}
  {\bibfnamefont {X.}~\bibnamefont {Ke}},\ }\bibfield  {title} {\bibinfo
  {title} {{Coexistence and Interaction of Spinons and Magnons in an
  Antiferromagnet with Alternating Antiferromagnetic and Ferromagnetic Quantum
  Spin Chains}},\ }\href {https://doi.org/10.1103/PhysRevLett.125.037204}
  {\bibfield  {journal} {\bibinfo  {journal} {Phys. Rev. Lett.}\ }\textbf
  {\bibinfo {volume} {125}},\ \bibinfo {pages} {037204} (\bibinfo {year}
  {2020})}\BibitemShut {NoStop}%
\bibitem [{\citenamefont {Gautreau}\ \emph {et~al.}(2021)\citenamefont
  {Gautreau}, \citenamefont {Saha},\ and\ \citenamefont
  {Birol}}]{Gautreau_PRM_2021_COHBdft}%
  \BibitemOpen
  \bibfield  {author} {\bibinfo {author} {\bibfnamefont {D.~M.}\ \bibnamefont
  {Gautreau}}, \bibinfo {author} {\bibfnamefont {A.}~\bibnamefont {Saha}},\
  and\ \bibinfo {author} {\bibfnamefont {T.}~\bibnamefont {Birol}},\ }\bibfield
   {title} {\bibinfo {title} {{First-principles characterization of the
  magnetic properties of ${\mathrm{Cu}}_{2}{(\mathrm{OH})}_{3}\mathrm{Br}$}},\
  }\href {https://doi.org/10.1103/PhysRevMaterials.5.024407} {\bibfield
  {journal} {\bibinfo  {journal} {Phys. Rev. Mater.}\ }\textbf {\bibinfo
  {volume} {5}},\ \bibinfo {pages} {024407} (\bibinfo {year}
  {2021})}\BibitemShut {NoStop}%
\bibitem [{\citenamefont {Xiao}\ \emph {et~al.}(2022)\citenamefont {Xiao},
  \citenamefont {Ouyang}, \citenamefont {Liu}, \citenamefont {Cao},
  \citenamefont {Wang},\ and\ \citenamefont
  {Tong}}]{XiaoOuyang_JPCM_2022_COHBangular}%
  \BibitemOpen
  \bibfield  {author} {\bibinfo {author} {\bibfnamefont {T.~T.}\ \bibnamefont
  {Xiao}}, \bibinfo {author} {\bibfnamefont {Z.~W.}\ \bibnamefont {Ouyang}},
  \bibinfo {author} {\bibfnamefont {X.~C.}\ \bibnamefont {Liu}}, \bibinfo
  {author} {\bibfnamefont {J.~J.}\ \bibnamefont {Cao}}, \bibinfo {author}
  {\bibfnamefont {Z.~X.}\ \bibnamefont {Wang}},\ and\ \bibinfo {author}
  {\bibfnamefont {W.}~\bibnamefont {Tong}},\ }\bibfield  {title} {\bibinfo
  {title} {{Angular dependence of spin-flop transition in triangular lattice
  antiferromagnet ${\mathrm{Cu}}_{2}{(\mathrm{OH})}_{3}\mathrm{Br}$}},\ }\href
  {https://doi.org/10.1088/1361-648X/ac69a0} {\bibfield  {journal} {\bibinfo
  {journal} {J. Phys.: Cond. Matter}\ }\textbf {\bibinfo {volume} {34}},\
  \bibinfo {pages} {275804} (\bibinfo {year} {2022})}\BibitemShut {NoStop}%
\bibitem [{\citenamefont {Zheng}\ \emph {et~al.}(2005)\citenamefont {Zheng},
  \citenamefont {Mori}, \citenamefont {Nishiyama}, \citenamefont {Higemoto},
  \citenamefont {Yamada}, \citenamefont {Nishikubo},\ and\ \citenamefont
  {Xu}}]{ZhengMori_PRB_2005_COHXstructure}%
  \BibitemOpen
  \bibfield  {author} {\bibinfo {author} {\bibfnamefont {X.~G.}\ \bibnamefont
  {Zheng}}, \bibinfo {author} {\bibfnamefont {T.}~\bibnamefont {Mori}},
  \bibinfo {author} {\bibfnamefont {K.}~\bibnamefont {Nishiyama}}, \bibinfo
  {author} {\bibfnamefont {W.}~\bibnamefont {Higemoto}}, \bibinfo {author}
  {\bibfnamefont {H.}~\bibnamefont {Yamada}}, \bibinfo {author} {\bibfnamefont
  {K.}~\bibnamefont {Nishikubo}},\ and\ \bibinfo {author} {\bibfnamefont
  {C.~N.}\ \bibnamefont {Xu}},\ }\bibfield  {title} {\bibinfo {title}
  {{Antiferromagnetic transitions in polymorphous minerals of the natural
  cuprates atacamite and botallackite
  ${\mathrm{Cu}}_{2}\mathrm{Cl}{(\mathrm{O}\mathrm{H})}_{3}$}},\ }\href
  {https://doi.org/10.1103/PhysRevB.71.174404} {\bibfield  {journal} {\bibinfo
  {journal} {Phys. Rev. B}\ }\textbf {\bibinfo {volume} {71}},\ \bibinfo
  {pages} {174404} (\bibinfo {year} {2005})}\BibitemShut {NoStop}%
\bibitem [{\citenamefont {Krivovichev}\ \emph {et~al.}(2017)\citenamefont
  {Krivovichev}, \citenamefont {Hawthorne},\ and\ \citenamefont
  {Williams}}]{Krivovichev_StrChem_2017_COHXstructure}%
  \BibitemOpen
  \bibfield  {author} {\bibinfo {author} {\bibfnamefont {S.~V.}\ \bibnamefont
  {Krivovichev}}, \bibinfo {author} {\bibfnamefont {F.~C.}\ \bibnamefont
  {Hawthorne}},\ and\ \bibinfo {author} {\bibfnamefont {P.~A.}\ \bibnamefont
  {Williams}},\ }\bibfield  {title} {\bibinfo {title} {{Structural complexity
  and crystallization: the Ostwald sequence of phases in the Cu$_2$(OH)$_3$Cl
  system (botallackite–atacamite–clinoatacamite)}},\ }\href
  {https://doi.org/10.1007/s11224-016-0792-z} {\bibfield  {journal} {\bibinfo
  {journal} {Struct. Chem.}\ }\textbf {\bibinfo {volume} {28}},\ \bibinfo
  {pages} {153} (\bibinfo {year} {2017})}\BibitemShut {NoStop}%
\bibitem [{\citenamefont {Skourski}\ \emph {et~al.}(2011)\citenamefont
  {Skourski}, \citenamefont {Kuz'min}, \citenamefont {Skokov}, \citenamefont
  {Andreev},\ and\ \citenamefont {Wosnitza}}]{SkourskiKuzmin_PRB_2011_PulseM}%
  \BibitemOpen
  \bibfield  {author} {\bibinfo {author} {\bibfnamefont {Y.}~\bibnamefont
  {Skourski}}, \bibinfo {author} {\bibfnamefont {M.~D.}\ \bibnamefont
  {Kuz'min}}, \bibinfo {author} {\bibfnamefont {K.~P.}\ \bibnamefont {Skokov}},
  \bibinfo {author} {\bibfnamefont {A.~V.}\ \bibnamefont {Andreev}},\ and\
  \bibinfo {author} {\bibfnamefont {J.}~\bibnamefont {Wosnitza}},\ }\bibfield
  {title} {\bibinfo {title} {{High-field magnetization of
  Ho${}_{2}$Fe${}_{17}$}},\ }\href {https://doi.org/10.1103/PhysRevB.83.214420}
  {\bibfield  {journal} {\bibinfo  {journal} {Phys. Rev. B}\ }\textbf {\bibinfo
  {volume} {83}},\ \bibinfo {pages} {214420} (\bibinfo {year}
  {2011})}\BibitemShut {NoStop}%
\bibitem [{\citenamefont {Clover}\ and\ \citenamefont
  {Wolf}(1970)}]{CloverWulf_RevSciInstr_1970_TDO}%
  \BibitemOpen
  \bibfield  {author} {\bibinfo {author} {\bibfnamefont {R.~B.}\ \bibnamefont
  {Clover}}\ and\ \bibinfo {author} {\bibfnamefont {W.~P.}\ \bibnamefont
  {Wolf}},\ }\bibfield  {title} {\bibinfo {title} {{Magnetic Susceptibility
  Measurements with a Tunnel Diode Oscillator}},\ }\href
  {https://doi.org/10.1063/1.1684598} {\bibfield  {journal} {\bibinfo
  {journal} {Rev. Sci. Instr.}\ }\textbf {\bibinfo {volume} {41}},\ \bibinfo
  {pages} {617} (\bibinfo {year} {1970})}\BibitemShut {NoStop}%
\bibitem [{\citenamefont {Ghannadzadeh}\ \emph {et~al.}(2013)\citenamefont
  {Ghannadzadeh}, \citenamefont {M\"oller}, \citenamefont {Goddard},
  \citenamefont {Lancaster}, \citenamefont {Xiao}, \citenamefont {Blundell},
  \citenamefont {Maisuradze}, \citenamefont {Khasanov}, \citenamefont {Manson},
  \citenamefont {Tozer}, \citenamefont {Graf},\ and\ \citenamefont
  {Schlueter}}]{GhannadzadehMoeller_PRB_2013_earlyTDO}%
  \BibitemOpen
  \bibfield  {author} {\bibinfo {author} {\bibfnamefont {S.}~\bibnamefont
  {Ghannadzadeh}}, \bibinfo {author} {\bibfnamefont {J.~S.}\ \bibnamefont
  {M\"oller}}, \bibinfo {author} {\bibfnamefont {P.~A.}\ \bibnamefont
  {Goddard}}, \bibinfo {author} {\bibfnamefont {T.}~\bibnamefont {Lancaster}},
  \bibinfo {author} {\bibfnamefont {F.}~\bibnamefont {Xiao}}, \bibinfo {author}
  {\bibfnamefont {S.~J.}\ \bibnamefont {Blundell}}, \bibinfo {author}
  {\bibfnamefont {A.}~\bibnamefont {Maisuradze}}, \bibinfo {author}
  {\bibfnamefont {R.}~\bibnamefont {Khasanov}}, \bibinfo {author}
  {\bibfnamefont {J.~L.}\ \bibnamefont {Manson}}, \bibinfo {author}
  {\bibfnamefont {S.~W.}\ \bibnamefont {Tozer}}, \bibinfo {author}
  {\bibfnamefont {D.}~\bibnamefont {Graf}},\ and\ \bibinfo {author}
  {\bibfnamefont {J.~A.}\ \bibnamefont {Schlueter}},\ }\bibfield  {title}
  {\bibinfo {title} {{Evolution of magnetic interactions in a pressure-induced
  Jahn-Teller driven magnetic dimensionality switch}},\ }\href
  {https://doi.org/10.1103/PhysRevB.87.241102} {\bibfield  {journal} {\bibinfo
  {journal} {Phys. Rev. B}\ }\textbf {\bibinfo {volume} {87}},\ \bibinfo
  {pages} {241102} (\bibinfo {year} {2013})}\BibitemShut {NoStop}%
\bibitem [{\citenamefont {Shi}\ \emph {et~al.}(2022)\citenamefont {Shi},
  \citenamefont {Dissanayake}, \citenamefont {Corboz}, \citenamefont
  {Steinhardt}, \citenamefont {Graf}, \citenamefont {Silevitch}, \citenamefont
  {Dabkowska}, \citenamefont {Rosenbaum}, \citenamefont {Mila},\ and\
  \citenamefont {Haravifard}}]{ShiDissanayakeCorboz_NatComm_2022_SCBOpressTDO}%
  \BibitemOpen
  \bibfield  {author} {\bibinfo {author} {\bibfnamefont {Z.}~\bibnamefont
  {Shi}}, \bibinfo {author} {\bibfnamefont {S.}~\bibnamefont {Dissanayake}},
  \bibinfo {author} {\bibfnamefont {P.}~\bibnamefont {Corboz}}, \bibinfo
  {author} {\bibfnamefont {W.}~\bibnamefont {Steinhardt}}, \bibinfo {author}
  {\bibfnamefont {D.}~\bibnamefont {Graf}}, \bibinfo {author} {\bibfnamefont
  {D.~M.}\ \bibnamefont {Silevitch}}, \bibinfo {author} {\bibfnamefont {H.~A.}\
  \bibnamefont {Dabkowska}}, \bibinfo {author} {\bibfnamefont {T.~F.}\
  \bibnamefont {Rosenbaum}}, \bibinfo {author} {\bibfnamefont {F.}~\bibnamefont
  {Mila}},\ and\ \bibinfo {author} {\bibfnamefont {S.}~\bibnamefont
  {Haravifard}},\ }\bibfield  {title} {\bibinfo {title} {{Discovery of quantum
  phases in the Shastry-Sutherland compound SrCu$_2$(BO$_3$)$_2$ under extreme
  conditions of field and pressure}},\ }\href
  {https://doi.org/10.1038/s41467-022-30036-w} {\bibfield  {journal} {\bibinfo
  {journal} {Nat. Commun.}\ }\textbf {\bibinfo {volume} {13}},\ \bibinfo
  {pages} {2301} (\bibinfo {year} {2022})}\BibitemShut {NoStop}%
\bibitem [{\citenamefont {Zvyagin}\ \emph {et~al.}(2004)\citenamefont
  {Zvyagin}, \citenamefont {Krzystek}, \citenamefont {{van Loosdrecht}},
  \citenamefont {Dhalenne},\ and\ \citenamefont
  {Revcolevschi}}]{Zvyagin_PhysB_2004_ESRinCuGeO3}%
  \BibitemOpen
  \bibfield  {author} {\bibinfo {author} {\bibfnamefont {S.~A.}\ \bibnamefont
  {Zvyagin}}, \bibinfo {author} {\bibfnamefont {J.}~\bibnamefont {Krzystek}},
  \bibinfo {author} {\bibfnamefont {P.~H.~M.}\ \bibnamefont {{van
  Loosdrecht}}}, \bibinfo {author} {\bibfnamefont {G.}~\bibnamefont
  {Dhalenne}},\ and\ \bibinfo {author} {\bibfnamefont {A.}~\bibnamefont
  {Revcolevschi}},\ }\bibfield  {title} {\bibinfo {title} {{High-field ESR
  study of the dimerized-incommensurate phase transition in the spin-Peierls
  compound CuGeO$_3$}},\ }\href
  {https://doi.org/https://doi.org/10.1016/j.physb.2004.01.009} {\bibfield
  {journal} {\bibinfo  {journal} {Physica B}\ }\textbf {\bibinfo {volume}
  {346-347}},\ \bibinfo {pages} {1} (\bibinfo {year} {2004})}\BibitemShut
  {NoStop}%
\bibitem [{\citenamefont
  {Dzyaloshinsky}(1958)}]{Dzyaloshinskii_JChemPhysSol_1958_DM}%
  \BibitemOpen
  \bibfield  {author} {\bibinfo {author} {\bibfnamefont {I.}~\bibnamefont
  {Dzyaloshinsky}},\ }\bibfield  {title} {\bibinfo {title} {{A thermodynamic
  theory of 'weak' ferromagnetism of antiferromagnetics}},\ }\href
  {https://doi.org/10.1016/0022-3697(58)90076-3} {\bibfield  {journal}
  {\bibinfo  {journal} {J. Phys. Chem. Solids}\ }\textbf {\bibinfo {volume}
  {4}},\ \bibinfo {pages} {241} (\bibinfo {year} {1958})}\BibitemShut {NoStop}%
\bibitem [{\citenamefont {Moriya}(1960)}]{Moriya_PR_1960_DM}%
  \BibitemOpen
  \bibfield  {author} {\bibinfo {author} {\bibfnamefont {T.}~\bibnamefont
  {Moriya}},\ }\bibfield  {title} {\bibinfo {title} {{Anisotropic Superexchange
  Interaction and Weak Ferromagnetism}},\ }\href
  {https://doi.org/10.1103/PhysRev.120.91} {\bibfield  {journal} {\bibinfo
  {journal} {Phys. Rev.}\ }\textbf {\bibinfo {volume} {120}},\ \bibinfo {pages}
  {91} (\bibinfo {year} {1960})}\BibitemShut {NoStop}%
\bibitem [{\citenamefont {Oshikawa}\ and\ \citenamefont
  {Affleck}(1997)}]{OshikawaAffleck_PRL_1997_staggered}%
  \BibitemOpen
  \bibfield  {author} {\bibinfo {author} {\bibfnamefont {M.}~\bibnamefont
  {Oshikawa}}\ and\ \bibinfo {author} {\bibfnamefont {I.}~\bibnamefont
  {Affleck}},\ }\bibfield  {title} {\bibinfo {title} {{Field-Induced Gap in
  $\mathit{S}\phantom{\rule{0ex}{0ex}}=\phantom{\rule{0ex}{0ex}}1/2$
  Antiferromagnetic Chains}},\ }\href
  {https://doi.org/10.1103/PhysRevLett.79.2883} {\bibfield  {journal} {\bibinfo
   {journal} {Phys. Rev. Lett.}\ }\textbf {\bibinfo {volume} {79}},\ \bibinfo
  {pages} {2883} (\bibinfo {year} {1997})}\BibitemShut {NoStop}%
\bibitem [{\citenamefont {Chen}\ \emph {et~al.}(2007)\citenamefont {Chen},
  \citenamefont {Stone}, \citenamefont {Kenzelmann}, \citenamefont {Batista},
  \citenamefont {Reich},\ and\ \citenamefont
  {Broholm}}]{ChenStone_PRB_2007_CDCstaggeredOP}%
  \BibitemOpen
  \bibfield  {author} {\bibinfo {author} {\bibfnamefont {Y.}~\bibnamefont
  {Chen}}, \bibinfo {author} {\bibfnamefont {M.~B.}\ \bibnamefont {Stone}},
  \bibinfo {author} {\bibfnamefont {M.}~\bibnamefont {Kenzelmann}}, \bibinfo
  {author} {\bibfnamefont {C.~D.}\ \bibnamefont {Batista}}, \bibinfo {author}
  {\bibfnamefont {D.~H.}\ \bibnamefont {Reich}},\ and\ \bibinfo {author}
  {\bibfnamefont {C.}~\bibnamefont {Broholm}},\ }\bibfield  {title} {\bibinfo
  {title} {{Phase diagram and spin Hamiltonian of weakly-coupled anisotropic
  $S=\frac{1}{2}$ chains in $\mathrm{Cu}{\mathrm{Cl}}_{2}\cdot
  2({(\mathrm{C}{\mathrm{D}}_{3})}_{2}\mathrm{S}\mathrm{O})$}},\ }\href
  {https://doi.org/10.1103/PhysRevB.75.214409} {\bibfield  {journal} {\bibinfo
  {journal} {Phys. Rev. B}\ }\textbf {\bibinfo {volume} {75}},\ \bibinfo
  {pages} {214409} (\bibinfo {year} {2007})}\BibitemShut {NoStop}%
\bibitem [{\citenamefont {Turov}\ \emph {et~al.}(2004)\citenamefont {Turov},
  \citenamefont {Kolchanov},\ and\ \citenamefont
  {Kurkin}}]{Turov_2004_AFMsbook}%
  \BibitemOpen
  \bibfield  {author} {\bibinfo {author} {\bibfnamefont {E.}~\bibnamefont
  {Turov}}, \bibinfo {author} {\bibfnamefont {A.}~\bibnamefont {Kolchanov}},\
  and\ \bibinfo {author} {\bibfnamefont {M.}~\bibnamefont {Kurkin}},\ }\href
  {https://search.worldcat.org/title/249527656} {\emph {\bibinfo {title}
  {{Symmetry and Physical Properties of Antiferromagnetics}}}}\ (\bibinfo
  {publisher} {Cambridge International Science Publishing},\ \bibinfo {year}
  {2004})\BibitemShut {NoStop}%
\bibitem [{\citenamefont {Lake}\ \emph {et~al.}(2005)\citenamefont {Lake},
  \citenamefont {Tennant}, \citenamefont {Frost},\ and\ \citenamefont
  {Nagler}}]{Lake_NatMat_2005_ChainScaling}%
  \BibitemOpen
  \bibfield  {author} {\bibinfo {author} {\bibfnamefont {B.}~\bibnamefont
  {Lake}}, \bibinfo {author} {\bibfnamefont {D.~A.}\ \bibnamefont {Tennant}},
  \bibinfo {author} {\bibfnamefont {C.~D.}\ \bibnamefont {Frost}},\ and\
  \bibinfo {author} {\bibfnamefont {S.~E.}\ \bibnamefont {Nagler}},\ }\bibfield
   {title} {\bibinfo {title} {{Quantum criticality and universal scaling of a
  quantum antiferromagnet.}},\ }\href {https://doi.org/10.1038/nmat1327}
  {\bibfield  {journal} {\bibinfo  {journal} {Nat. Mater.}\ }\textbf {\bibinfo
  {volume} {4}},\ \bibinfo {pages} {329} (\bibinfo {year} {2005})}\BibitemShut
  {NoStop}%
\bibitem [{\citenamefont {Coldea}\ \emph {et~al.}(2003)\citenamefont {Coldea},
  \citenamefont {Tennant},\ and\ \citenamefont
  {Tylczynski}}]{ColdeaTennant_PRB_2003_Cs2CuCl4continua}%
  \BibitemOpen
  \bibfield  {author} {\bibinfo {author} {\bibfnamefont {R.}~\bibnamefont
  {Coldea}}, \bibinfo {author} {\bibfnamefont {D.~A.}\ \bibnamefont
  {Tennant}},\ and\ \bibinfo {author} {\bibfnamefont {Z.}~\bibnamefont
  {Tylczynski}},\ }\bibfield  {title} {\bibinfo {title} {{Extended scattering
  continua characteristic of spin fractionalization in the two-dimensional
  frustrated quantum magnet ${\mathrm{Cs}}_{2}{\mathrm{CuCl}}_{4}$ observed by
  neutron scattering}},\ }\href {https://doi.org/10.1103/PhysRevB.68.134424}
  {\bibfield  {journal} {\bibinfo  {journal} {Phys. Rev. B}\ }\textbf {\bibinfo
  {volume} {68}},\ \bibinfo {pages} {134424} (\bibinfo {year}
  {2003})}\BibitemShut {NoStop}%
\bibitem [{\citenamefont {Smirnov}\ \emph {et~al.}(2012)\citenamefont
  {Smirnov}, \citenamefont {Povarov}, \citenamefont {Petrov},\ and\
  \citenamefont {Shapiro}}]{SmirnovPovarov_PRB_2012_ESRordered}%
  \BibitemOpen
  \bibfield  {author} {\bibinfo {author} {\bibfnamefont {A.~I.}\ \bibnamefont
  {Smirnov}}, \bibinfo {author} {\bibfnamefont {{\relax K. Yu}.}~\bibnamefont
  {Povarov}}, \bibinfo {author} {\bibfnamefont {S.~V.}\ \bibnamefont
  {Petrov}},\ and\ \bibinfo {author} {\bibfnamefont {{\relax A.
  Ya}.}~\bibnamefont {Shapiro}},\ }\bibfield  {title} {\bibinfo {title}
  {{Magnetic resonance in the ordered phases of the two-dimensional frustrated
  quantum magnet Cs${}_{2}$CuCl${}_{4}$}},\ }\href
  {https://doi.org/10.1103/PhysRevB.85.184423} {\bibfield  {journal} {\bibinfo
  {journal} {Phys. Rev. B}\ }\textbf {\bibinfo {volume} {85}},\ \bibinfo
  {pages} {184423} (\bibinfo {year} {2012})}\BibitemShut {NoStop}%
\bibitem [{\citenamefont {Nawa}\ \emph {et~al.}(2020)\citenamefont {Nawa},
  \citenamefont {Hirai}, \citenamefont {Kofu}, \citenamefont {Nakajima},
  \citenamefont {Murasaki}, \citenamefont {Kogane}, \citenamefont {Kimata},
  \citenamefont {Nojiri}, \citenamefont {Hiroi},\ and\ \citenamefont
  {Sato}}]{NawaHirai_PRR_2020_CROCspinons}%
  \BibitemOpen
  \bibfield  {author} {\bibinfo {author} {\bibfnamefont {K.}~\bibnamefont
  {Nawa}}, \bibinfo {author} {\bibfnamefont {D.}~\bibnamefont {Hirai}},
  \bibinfo {author} {\bibfnamefont {M.}~\bibnamefont {Kofu}}, \bibinfo {author}
  {\bibfnamefont {K.}~\bibnamefont {Nakajima}}, \bibinfo {author}
  {\bibfnamefont {R.}~\bibnamefont {Murasaki}}, \bibinfo {author}
  {\bibfnamefont {S.}~\bibnamefont {Kogane}}, \bibinfo {author} {\bibfnamefont
  {M.}~\bibnamefont {Kimata}}, \bibinfo {author} {\bibfnamefont
  {H.}~\bibnamefont {Nojiri}}, \bibinfo {author} {\bibfnamefont
  {Z.}~\bibnamefont {Hiroi}},\ and\ \bibinfo {author} {\bibfnamefont {T.~J.}\
  \bibnamefont {Sato}},\ }\bibfield  {title} {\bibinfo {title} {{Bound spinon
  excitations in the spin-$\frac{1}{2}$ anisotropic triangular antiferromagnet
  ${\mathrm{Ca}}_{3}\mathrm{Re}{\mathrm{O}}_{5}{\mathrm{Cl}}_{2}$}},\ }\href
  {https://doi.org/10.1103/PhysRevResearch.2.043121} {\bibfield  {journal}
  {\bibinfo  {journal} {Phys. Rev. Research}\ }\textbf {\bibinfo {volume}
  {2}},\ \bibinfo {pages} {043121} (\bibinfo {year} {2020})}\BibitemShut
  {NoStop}%
\bibitem [{\citenamefont {Gurevich}\ and\ \citenamefont
  {Melkov}(1996)}]{GurevichMelkov_1996_Wavesbook}%
  \BibitemOpen
  \bibfield  {author} {\bibinfo {author} {\bibfnamefont {A.~G.}\ \bibnamefont
  {Gurevich}}\ and\ \bibinfo {author} {\bibfnamefont {G.~A.}\ \bibnamefont
  {Melkov}},\ }\href {https://doi.org/10.1201/9780138748487} {\emph {\bibinfo
  {title} {{Magnetization Oscillations and Waves}}}}\ (\bibinfo  {publisher}
  {CRC Press, U.K.},\ \bibinfo {year} {1996})\BibitemShut {NoStop}%
\bibitem [{\citenamefont {Reinold}\ \emph {et~al.}(2024)\citenamefont
  {Reinold}, \citenamefont {Berger}, \citenamefont {Raczkowski}, \citenamefont
  {Zhao}, \citenamefont {Kohama}, \citenamefont {Gen}, \citenamefont
  {Gorbunov}, \citenamefont {Skourski}, \citenamefont {Zherlitsyn},
  \citenamefont {Assaad}, \citenamefont {Lorenz},\ and\ \citenamefont
  {Wang}}]{ReinoldBerger_arXiv_2024_COHBHiib}%
  \BibitemOpen
  \bibfield  {author} {\bibinfo {author} {\bibfnamefont {A.}~\bibnamefont
  {Reinold}}, \bibinfo {author} {\bibfnamefont {L.}~\bibnamefont {Berger}},
  \bibinfo {author} {\bibfnamefont {M.}~\bibnamefont {Raczkowski}}, \bibinfo
  {author} {\bibfnamefont {Z.}~\bibnamefont {Zhao}}, \bibinfo {author}
  {\bibfnamefont {Y.}~\bibnamefont {Kohama}}, \bibinfo {author} {\bibfnamefont
  {M.}~\bibnamefont {Gen}}, \bibinfo {author} {\bibfnamefont {D.~I.}\
  \bibnamefont {Gorbunov}}, \bibinfo {author} {\bibfnamefont {Y.}~\bibnamefont
  {Skourski}}, \bibinfo {author} {\bibfnamefont {S.}~\bibnamefont
  {Zherlitsyn}}, \bibinfo {author} {\bibfnamefont {F.~F.}\ \bibnamefont
  {Assaad}}, \bibinfo {author} {\bibfnamefont {T.}~\bibnamefont {Lorenz}},\
  and\ \bibinfo {author} {\bibfnamefont {Z.}~\bibnamefont {Wang}},\ }\bibfield
  {title} {\bibinfo {title} {{Magnetization Process of a Quasi-Two-Dimensional
  Quantum Magnet: Two-Step Symmetry Restoration and Dimensional Reduction}},\
  }\href {http://arxiv.org/abs/2411.09541} {\bibfield  {journal} {\bibinfo
  {journal} {arXiv}\ ,\ \bibinfo {pages} {2411.09541}} (\bibinfo {year}
  {2024})}\BibitemShut {NoStop}%
\bibitem [{\citenamefont {Toth}\ and\ \citenamefont
  {Lake}(2015)}]{TothLake_JPCM_2015_SpinW}%
  \BibitemOpen
  \bibfield  {author} {\bibinfo {author} {\bibfnamefont {S.}~\bibnamefont
  {Toth}}\ and\ \bibinfo {author} {\bibfnamefont {B.}~\bibnamefont {Lake}},\
  }\bibfield  {title} {\bibinfo {title} {{Linear spin wave theory for
  single-$Q$ incommensurate magnetic structures}},\ }\href
  {https://doi.org/10.1088/0953-8984/27/16/166002} {\bibfield  {journal}
  {\bibinfo  {journal} {J. Phys.: Condens. Matter}\ }\textbf {\bibinfo {volume}
  {27}},\ \bibinfo {pages} {166002} (\bibinfo {year} {2015})}\BibitemShut
  {NoStop}%
\end{thebibliography}%
\end{document}